\documentclass[aps,pre,reprint,superscriptaddress]{revtex4-1} 
\usepackage{graphicx}
\usepackage{color}
\usepackage{amsmath}
\usepackage{paralist}
\usepackage{indentfirst}  
\usepackage{bm}    
\usepackage{amsfonts}
\usepackage{tikz}
\newcommand{\kinectTM}{Kinect\texttrademark}
\newcommand{\kinectTMS}{\kinectTM~}
\newcommand{\Eavg}{\mathbb{E}}  
\newcommand{\intGrI}{\mathcal{I}}
\newcommand{\ensavg}[1]{\langle #1 \rangle_{\mbox{\footnotesize \texttt ens}}}

\begin{document}


\title{Physics-based modeling and data representation of pedestrian pairwise interactions}


\author{Alessandro Corbetta}
\affiliation{Department of Applied Physics, Eindhoven University of Technology,  5600 MB Eindhoven, The Netherlands}

\author{Jasper A. Meeusen}
\affiliation{Department of Applied Physics, Eindhoven University of Technology,  5600 MB Eindhoven, The Netherlands}

\author{Chung-min Lee}
\affiliation{Department of Mathematics and Statistics, California State University Long Beach,  90840, Long Beach, CA, USA}

\author{Roberto Benzi}
\affiliation{Department of Physics and INFN, University of Tor Vergata
, I-00133 Rome, Italy}

\author{Federico Toschi}
\affiliation{Department of Applied Physics, Department of Mathematics and Computer Science, Eindhoven University of
Technology - 5600 MB Eindhoven, The Netherlands and CNR-IAC, I-00185 Rome, Italy}



\date{\today}

\begin{abstract}

The possibility to understand and to quantitatively model the physics
of the interactions between pedestrians walking in crowds has compelling
relevant applications, e.g. related to the efficient design and the
safety of civil infrastructures. In this work we study
pedestrian-pedestrian interactions from observational experimental
data in diluted pedestrian crowds. While in motion, pedestrians continuously adapt
their walking paths trying to preserve mutual comfort distances and to
avoid collisions. In mathematical models this behavior is typically
modeled via ``social'' interaction forces. 

Leveraging on a
high-quality, high-statistics dataset -- composed of few millions of
real-life trajectories acquired from state of the art observational experiments (about 6 months of high-resolution pedestrian tracks acquired in a train station)
--
we develop a quantitative model capable of
addressing  interactions in the case of binary collision
avoidance. We model interactions in terms of both long-range (sight
based) and short-range (hard-contact avoidance) forces, which we superimpose 
to  our Langevin model for non-interacting
pedestrian motion [Corbetta et al. Phys.\,Rev.\,E 95, 032316, 2017] (here further tested and extended).
The new model that we propose here features a Langevin dynamics with ``fast'' random
velocity fluctuations that are superimposed to the ``slow'' dynamics of a hidden model variable: the
``intended'' walking path. In case of interactions, social forces may act both on the intended path and on the actual walked path.
The model is capable of reproducing quantitatively relevant statistics of the collision avoidance motion, such as the statistics of the  side displacement and of the passing speed. Rare occurrences of actual bumping events are also recovered.

Furthermore, comparing with large datasets of real-life tracks 
involves an additional computational challenge so far neglected: identifying automatically,
within a database containing very heterogeneous conditions, only the relevant events
corresponding to binary avoidance interactions. In order to tackle this challenge, we propose a novel and general approach
based on a graph representation of pedestrian trajectories, which allows
us to effectively operate complexity reduction for efficient data classification and selection.
\end{abstract}


\maketitle

\section{Introduction}\label{sect:intro}

When we walk in a known environment or we explore a new venue, a path
is planned in our minds (our  ``intended path''). As other pedestrians approach us, or as we
learn features of the environment (e.g. better directions toward a target destination),
this path is continuously adjusted. Either as an impulsive act
or as a timely planned adjustment, we side-step to maintain comfort
distances among ourselves and other close by pedestrians. This comes
with a modification of our intended walking paths that ``bend'' in order to prevent
contacts or collisions with others.

\begin{figure}[!t!]
  \centering
  \includegraphics[width=.45\textwidth, trim={.5cm .5cm .5cm .5cm},clip]{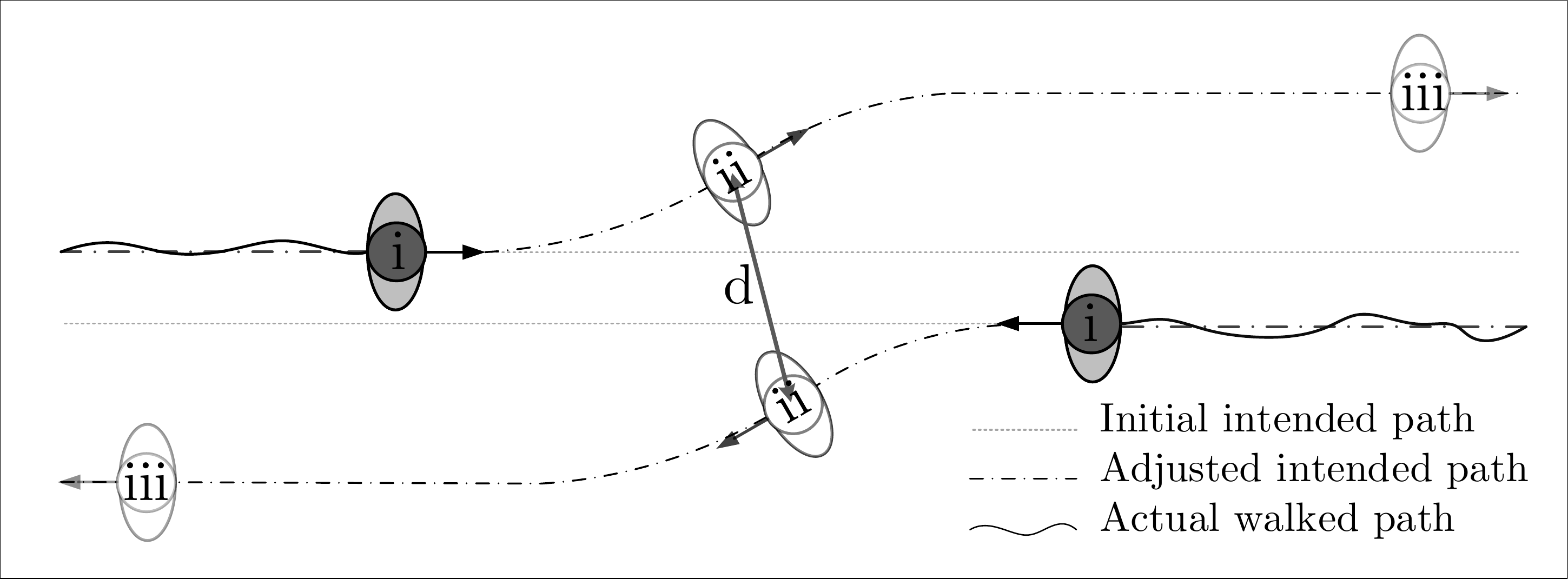}
  \caption{Mutual avoidance  of two pedestrians walking in
    opposite directions (``pairwise avoidance''). 
    Pedestrians walk trying to follow straight ``intended'' paths (snapshot i), around which
    they perform random fluctuations (cf. Sect.~\ref{sect:diluted}). 
    Individual motions however remain influenced by the dynamics of peers.  
    As a peer approaches, the intended path is adjusted (snapshot ii) to ensure maintenance of  mutual comfort distance (snapshot iii). We investigate
    and model quantitatively 
    the avoidance dynamics (cf. Sect.~\ref{sec:interactions}) with
    reference to three distances ($d$), transversal to the motion,  and characteristic  of the
    interaction: (i) before adjusting the intended path (at the entrance of our observation window),
    at the moment of side-by-side (ii) and when pedestrians leave our observation window (iii).}
\label{fig:sketch-Trajectories}
\end{figure}

The aim of this paper is to quantitatively
understand and model the dynamics behind these path changes -- in the
simplest condition of two pedestrians walking in opposite directions,
trying to avoid each other (``pairwise avoidance'' - 
cf. Fig.~\ref{fig:sketch-Trajectories}).
This is the 
first necessary step  to understand the interaction physics
between pedestrians, before attempting to tackle more complex situations.

The dynamics of  path changes is a challenging subject within the
broader and compelling issue of understanding the flow of pedestrian
crowds~\cite{cristiani2014BOOK}. This scientific topic is not only
fascinating, because of its connections with the physics of emerging
complexity~\cite{Mou-pnas}, pattern formation~\cite{moussaid2009collective,Moussadrspb.2009.0405} and active
matters~\cite{Lutz,bellomo2012modeling}, but it is also extremely
relevant for its applications for the design, safety and 
performance of civil facilities~\cite{Sey1,Duives}.

Because of the macroscopic analogies between crowd and fluid flows~\cite{hughes2003flow},
modeling pedestrian dynamics in terms of interacting ``matter''
particles has become an emerging approach~\cite{bellomo2012modeling}. This analogy underlies
proper translations between \textit{passive} fluid particles -- that
move under the action of classic interaction forces -- and
\textit{active} pedestrians in crowds that interact via
``social'' forces~\cite{helbing1995PRE}. Social forces abstract pedestrian-pedestrian
interactions in a Newtonian-like way. As such, we expect that mutual
repulsive interaction (social) forces may act for ensuring comfort distances
and collision avoidance,  possibly modifying pedestrians' intended paths.

\begin{figure}[!ht]
 \centering
\includegraphics[width=.45\textwidth]{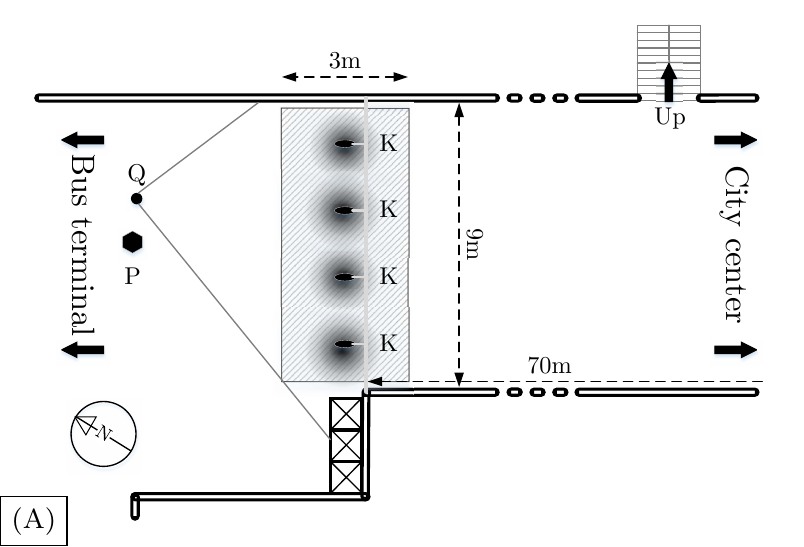}
\includegraphics[width=.45\textwidth]{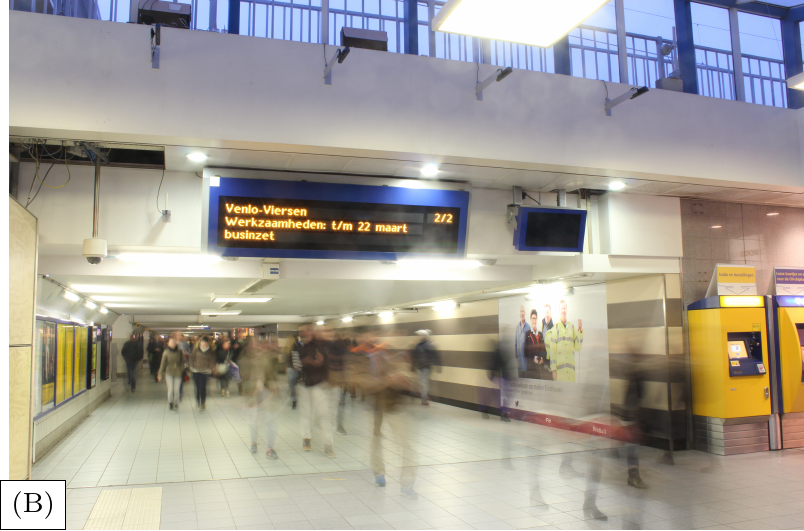}
  \caption{Experimental setup used in our $6$ months-long pedestrian tracking campaign at the train station in Eindhoven (NL); cf. Sect.~\ref{sect:campaign}.
    (A) Planar layout of the north entrance hall of Eindhoven
    train station (as it was between 2013 and early 2015). The measurement
    area is shaded. The entrance hall, facing the bus terminal of the
    city, leads to a $70\,$m long tunnel connecting to the south side
    of the city. The rail platforms are reachable from the tunnel via
    several side staircases. Our pedestrian dynamics recordings
    employed four overhead \kinectTMS~\cite{Kinect} sensors (\textit{K}) with
    partially overlapping view. The sensors were attached to the side of the overhang and are supported by metallic arms (cf. (B)). The snapshot
    (B) was taken from point Q, i.e. in the vicinity of a
    structural pillar (P) (about $5\,$m upstream of the recording
    area).} 
\label{fig:sketch-Eindhoven_Setup}
\end{figure}

Despite the growing scientific and technological interest for the
motion of pedestrian crowds, our quantitative understanding remains
relatively limited, especially in comparison to other kind of
``active matter systems''~\cite{RevModPhys.85.1143}. 
A major limitation comes from the fact that
high-quality experimental data, with high resolution in space and time,
still remain scarce.  An important point to be understood is
that pedestrian motion has a strong variability, which can 
be decoupled from average trends only by considering measurements with extremely
high statistics. For instance, in the case of a narrow
corridor~\cite{corbetta2016fluctuations}, one needs tens
of thousands of measured trajectories to estimate the amplitude of the observables'
fluctuations (e.g. fluctuations in walking position, velocity, etc.) 
and to characterize the occurrence of  related rare events. 
In this paper, we employ data from tens of thousands  of avoidance events to investigate quantitatively and model the changes in intended paths from pairwise avoidance. 
The measurement of these events was achieved through a months-long real-life experimental campaign that we performed
in the main walkway of the train station of Eindhoven, the Netherlands, with state-of-the-art automated pedestrian tracking (see Fig.~\ref{fig:sketch-Eindhoven_Setup}). 
To the best of our knowledge, such an investigation has never been carried out before with the
accuracy and statistics as reported in this work.

The current scarceness of high-quality measurement data is probably
related to technical challenges connected to the acquisition of
pedestrian trajectory data. Collecting data in real-life conditions
demands robust individual tracking techniques, i.e. that remain
accurate regardless of factors such as illumination, clothing, presence of
objects, crowd density, and so on.

The analysis of pedestrian dynamics 
in a real-life setting involves, moreover, an additional challenge so far neglected: the
automated crowd scenario classification.
To illustrate this, let us consider the trajectories collected over weeks in a measurement zone within, e.g., a station or a mall.
These trajectories will certainly encompass different and alternating crowd scenarios. For instance, these can include:
pedestrians walking undisturbed (i.e. with no peers walking in their neighborhood),  pedestrians in small or large social groups, 
diluted or dense crowd streams in counter-flows, diluted or dense crowd streams in  co-flows, and so on (indeed all these scenarios occur in the measurements considered in this paper, cf. Fig.~\ref{fig:sketch-Eindhoven_Setup}).
If we focus on a given scenario that is defined by a set of parameters (for instance, pedestrians walking in a uni-directional flow at an assigned density level), and we compare the measurements from the occurrences of such a scenario, we expect to observe analogous features modulo random fluctuations. Besides, as the number of observed occurrences increases, we can quantify with higher and higher accuracy the statistics of fluctuations and of the rare events characteristic of the scenario.
For instance, people walking undisturbed are expected to have similar speed within fluctuations. If, instead, we consider social groups, we expect to measure velocities consistently lower than in the undisturbed case~\cite{10.1371/journal.pone.0010047}. 
Similarly, we expect counter-flow occurrences to exhibit mutual similarities, yet to feature different characteristic fluxes than co-flow conditions~\cite{corbetta2016continuous,kretz2006experimental}.
The scenario considered in this paper involves pairs of pedestrians mutually avoiding each other: to analyze the scenario including statistic fluctuations, we aggregate and analyze as ensemble
 all the trajectories occurred under such conditions.
When focusing on a specific scenario and conducting our investigation based on  the scenario's occurrences (in the following also referred to as realizations) we are performing  a ``virtual experiment''.  
In such a virtual experiment we analyze a subcollection of the whole experimental dataset pertaining to the given target scenario, which it is itself defined by a set of  control parameters (e.g. number of pedestrians involved, flow conditions considered, etc.). %
All occurrences of such scenario in the experimental dataset constitute the virtual experiment data, which we then explore and study as in a traditional laboratory experiment. 
Differently from a laboratory experiment, the pedestrians involved are not instructed to perform a pre-defined dynamics (cf. e.g.~\cite{6957746}), rather they can freely walk, without potential biases from the experimental setting.  
Identifying the subset of trajectories belonging to a target scenario  is thus a necessary first step in our investigation.
Since we deal with hundreds of thousands of trajectories, this identification cannot be performed manually.
In fact, this would demand an exhaustive visual analysis of thousands of hours of sensors' footage (something that has been routinely performed by humans in other smaller scales investigations, e.g.  to select groups 
in~\cite{PhysRevE.89.012811}, to classify walking patterns
in~\cite{Tamura2013}, or to isolate people waiting
in~\cite{seitzTGF15}). More in general, this identification task underlies a classification problem in which we associate each trajectory to its scenario.

Automatizing the trajectory classification task is the second aim of this paper, and it is instrumental to analyze the dynamics of path changes.
While automatic classification is widely studied in connection e.g. with images, text or speech content~\cite{kotsiantis2007supervised}, 
to the best of our knowledge this topic remains yet not addressed in the context of scenarios made of (pedestrian) trajectories.
Once more, this  likely relates with the fact that extensive data collection 
campaigns for pedestrian dynamics remain a rarity.

In this paper we target a two-fold state-of-the-art advancement. 
First,  we propose a novel representation strategy for pedestrian dynamics measurements,
based on graphs, to formally identify scenarios and
 automatically classify and select 
real-life trajectory data on such basis. 
Second,  we address quantitatively the dynamics of path changes and related pedestrian-pedestrian 
social forces in case of avoidance events involving two individuals 
(i.e. no third individual plays a role in the dynamics). 
For this, we propose a Langevin-like model, built extending our
previous quantitative model for the diluted (i.e. undisturbed/non-interacting) pedestrian dynamics~\cite{corbetta2016fluctuations} 
(cf., e.g.,~\cite{Lutz,coffey2004applications} for a general modeling reference on Langevin equations). 
The model is constructed in two steps: first, we generalize the diluted dynamics model to address a richer phenomenology, 
which is given by a mixture of pedestrians walking and (in tiny percentages) running. 
Second, we introduce and validate pairwise social forces that act simultaneously on the actual trajectory and on the intended path 
which we consider too a model variable. 
This force model enables us to  reproduce quantitatively our measurements of the pairwise avoidance dynamics including fluctuations and rare events (actual impacts).

\noindent The content of this paper is as follows: in Sect.~\ref{sect:campaign}, we
describe our measurement campaign and the acquired data, in the order of millions of trajectories.
This sets a basis for both the methodological
and the modeling contributions of this work, that are, respectively, in Sect.~\ref{sect:graph} and in Sect.~\ref{sect:diluted}-\ref{sec:interactions}.
In the methodological Section~\ref{sect:graph} we tackle the trajectory selection and classification issue.
In Sect.~\ref{sect:diluted} we address the motion of undisturbed pedestrians.
This is a necessary building block for Sect.~\ref{sec:interactions} in which we analyze 
and model the dynamics of 
pairwise avoidance and of the intended path.
A final discussion closes the paper.

\section{Measurement campaign}\label{sect:campaign}
The pedestrian dynamics data employed in this work 
 have been collected in the
period October 2014 -- March 2015 through a 24/7 real-life campaign at
Eindhoven train station. Our data acquisition took place in the
initial section of the main walkway of the station as presented
in Fig.~\ref{fig:sketch-Eindhoven_Setup}. The walkway is one of the
major pedestrian pathways between the north side and the south side
of the city with crowd traffic during the entire day. Different
dynamics ordinarily occur, such as co-flows and counter-flows with
density ranging from extremely low (one pedestrian in the entire walkway at night time)
to high during the morning peak commute times~\cite{corbetta2016continuous}.
We aimed at an
exhaustive individual tracking with high space- and time-resolution and, overall,
we collected about $100\,000$ trajectories per day and approximately
$5$ millions in total.

Four overhead Microsoft \kinectTMS sensors~\cite{Kinect} with partially overlapping
view recorded imaging-like data, specifically \textit{depth maps}, at
the rate of $15$ frames per second. Depth maps encode in gray-scale
levels the distance between each filmed pixel and the camera
plane; thus regions closer to the overhead sensors, such as heads,
result in darker shades. We blend the four depth map signals into a
single stream covering the entire measurement region 
of which we report few frames (already post-processed to
include e.g. individual trajectories) in
Fig.~\ref{fig:sketch-graph-depth-maps}. As in our previous
investigations 
(e.g.~\cite{corbetta2014TRP,corbettaTGF15}) and following the 
articles~\cite{seer2014kinects,brscic2013person}, we use cluster-based
analyses of depth maps to perform accurate localization of pedestrians
bodies and heads on a frame-by-frame basis. Finally, we employ particle-tracking
algorithms to extract individual trajectories from the output
localization step. We leave further technical details on the
detection and tracking procedures to Appendix~\ref{app:tracking}.

\begin{figure}[!ht]
\centering
\includegraphics[width=.45\textwidth, trim={.5cm .5cm .5cm .5cm},clip]{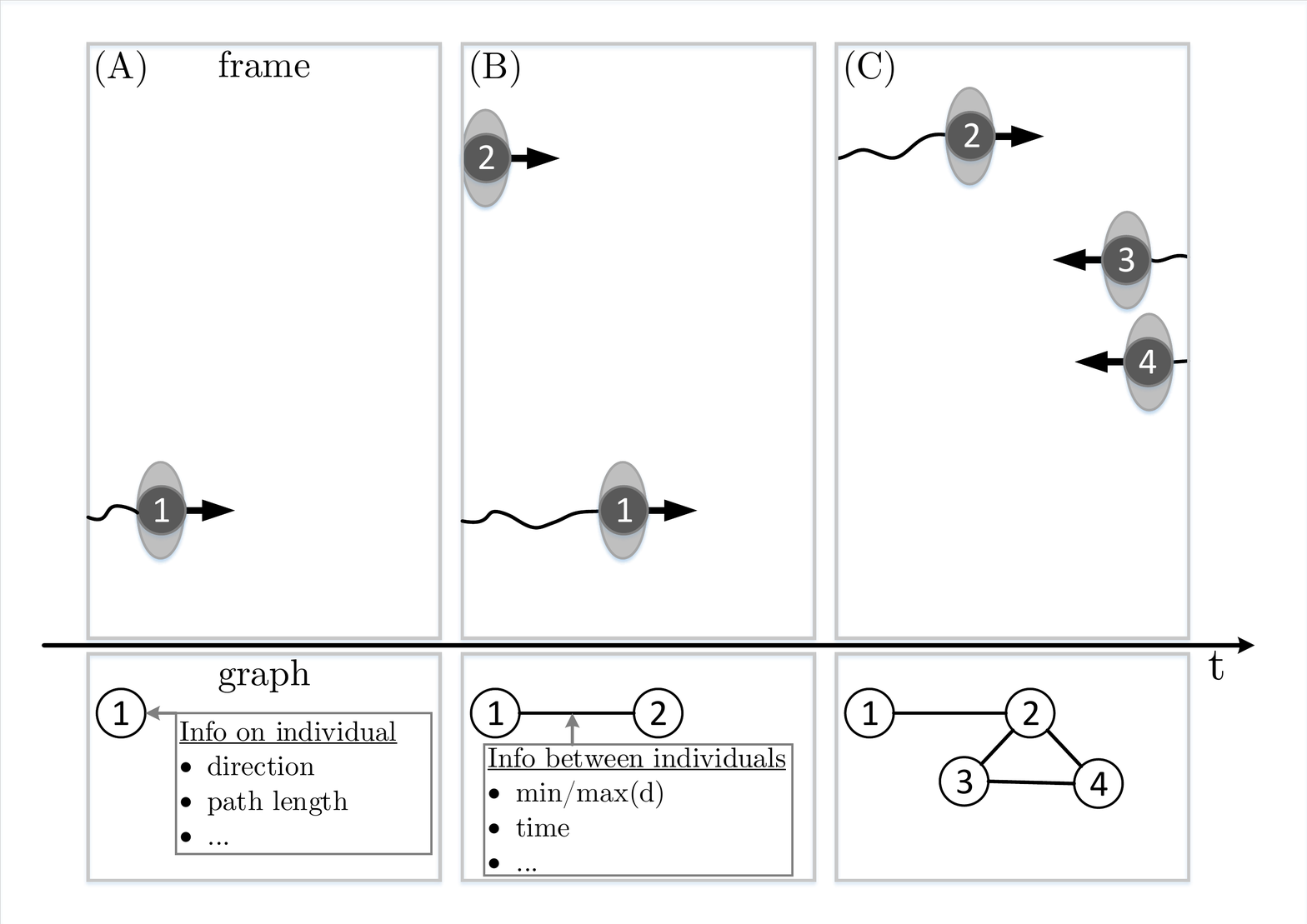}
\caption{We represent the recorded crowd dynamics with a graph $G$. This
  reduced description enables automatic classification of different
  flow scenarios (cf. Sect.~\ref{sect:graph}). Employing the three sample frames, (A), (B), (C), we
  schematize the graph construction algorithm. As a pedestrian,
  e.g. pedestrian ``$1$'' in (A), appears in our recording window, we add
  a corresponding node in the graph $G$. When two pedestrians are
  simultaneously in the recording window, e.g. ``$1$'' and ``$2$'' in
  (B), we connect the associated nodes with an edge. As further
  pedestrians are recorded the graph is expanded. In (C), we imagine that 
  pedestrians ``$3$'' and ``$4$'' entered the observation window after pedestrian
  ``$1$'' left. Therefore, their nodes are connected to one another and
  further just with the node representing pedestrian ``$2$''. We include additional information in the
  graph, crucial for the classification task: each node is annotated
  with scalar observables of the associated pedestrian trajectory
  (e.g. average velocity, direction) and each edge is weighted with
  scalar observables of the pairwise dynamics of the nodes
  (e.g. minimum and maximum distance, joint observation time).  }
\label{fig:sketch-Graph_creation}
\end{figure}

\begin{table}
  \caption{Bijective correspondences between real-life measurements and their representation in the graph $G$: summary of concepts and notation (cf. Sect.~\ref{sect:graph}). \label{tab:grap-symbols}}
  \begin{ruledtabular}
    \begin{tabular}{p{.22\textwidth} p{.22\textwidth} }
      Real-life measurement & Graph representation \\
      \hline\hline
      Trajectory set $\Gamma$ & Graph $G$ \\
      \hline
      Trajectory (i.e. pedestrian) \newline $p\in\Gamma$ & Node \newline $p\in G$ \\
      \hline
      $p$ and $q$ \newline interacting pedestrians  & edge $e=(p,q)$ in $\tilde{G}$ \\
      \hline
      Scenario  & Set of conditions identifying a sub-graph $\tilde{G}_s \subset \tilde{G}$ \\
      \hline
      Realizations of the scenario & Connected components in $\tilde{G}_s$ \\
      \hline
      Pedestrians walking undisturbed (all realizations) & Sub-graph $\tilde{G}_{1}\subset \tilde{G}$ \newline of the singleton nodes\\
      \hline
      Undisturbed pedestrian $p$ (one realization) & $p\in\tilde{G}_{1}$ \\
      \hline
      Pedestrians in pairwise avoidance (all realizations) &  Sub-graph $\tilde{G}_{2,a}\subset \tilde{G}$ \newline of dyads with opposite walking direction \\ 
      \hline
      Single pairwise avoidance event of $p$ and $q$ \newline (one realization) & Dyad $\{p,q\}$ in  $\tilde{G}_{2,a}$ \newline connected by an edge       
    \end{tabular}
  \end{ruledtabular}
\end{table}

\section{Representation and classification of crowd flow data}\label{sect:graph}

In this section we define two subsets of the trajectories collected in our measurement campaign (cf. Sect.~\ref{sect:campaign}), 
which we will use to investigate, respectively, the dynamics of undisturbed flows, considered in Sect.~\ref{sect:diluted}, 
and the dynamics of the pairwise avoidance, considered in Sect.~\ref{sec:interactions}. 
These subsets will be the output of a more general representation and classification construct, based on graphs, here introduced.

The underlying issue, as stated in Sect.~\ref{sect:intro}, is that large-scale measurements
 of pedestrian dynamics in real-life conditions
typically include different scenarios frequently and randomly
changing (see also~\cite{corbetta-colldyn17}).
For instance, around commuting time, in the walkway
in Fig.~\ref{fig:sketch-Eindhoven_Setup}, the flow changes abruptly
from diluted to dense. Every few seconds, the typical bi-directional
pedestrian flow  rapidly turns into a uni-directional dense stream
composed of the passengers just arrived by train.
The scenarios of our interest (undisturbed flow, pairwise avoidance) happen as well, yet  
alternating at random with others. Instead, we would like to perform ``virtual experiments'' 
investigating the dynamics of these scenarios one at a time, 
and use the data from the realizations of individual scenarios to increment our statistics. 
For instance, thousands of times per day, pedestrians walking 
undisturbed cross our measurement area. We expect all of them to
exhibit a similar behavior whose statistics we can accurately determine thanks to the large number of trajectories.  Analogously, we expect pairs of pedestrians in avoidance
(cf. Fig.~\ref{fig:sketch-Trajectories}) to show similar features as all realizations of the same dynamics.

In the next subsections we provide a strategy to  perform  virtual experiments. This involves the capability of 
\begin{inparaenum}
\item defining formally and quantitatively scenarios, and 
\item efficiently classifying and aggregating trajectories based on whether or not they are realizations of such scenarios.
\end{inparaenum}

In conceptual terms, given the set of measured trajectories, say the set $\Gamma = \{p\}$, we construct 
a representation of $\Gamma$ in terms of a graph $G$ with a bijective correspondence between trajectories and graph nodes.
This representation, reduced in complexity with respect to the original dataset,  suitably allows us to define scenarios
as conditions that identify sub-graphs of $G$ (Sect.~\ref{sect:representation}).
The desired output, i.e. sets of trajectories that are realization of these scenarios, are associated to the connected components of these sub-graphs. 
Trajectories occurring in the same instance of a scenario are in the same connected component 
(and, conversely, the complement of the sub-graph identify trajectories that are not realization of the given virtual experiment).

In Sect.~\ref{sect:classification} we define two sub-graphs $\tilde{G}_1 \subset G$ and $\tilde{G}_{2,a}\subset G$, whose connected components identify, respectively, realizations of diluted flows  and of pairwise avoidance, that we will use as experimental comparisons for the models considered respectively in Sect.~\ref{sect:diluted} and Sect.~\ref{sec:interactions}. In Tab.~\ref{tab:grap-symbols} we report a summary of the symbols and concepts used throughout this section.

\begin{figure}[ht!]
  \centering
  \includegraphics[width=0.45\textwidth]{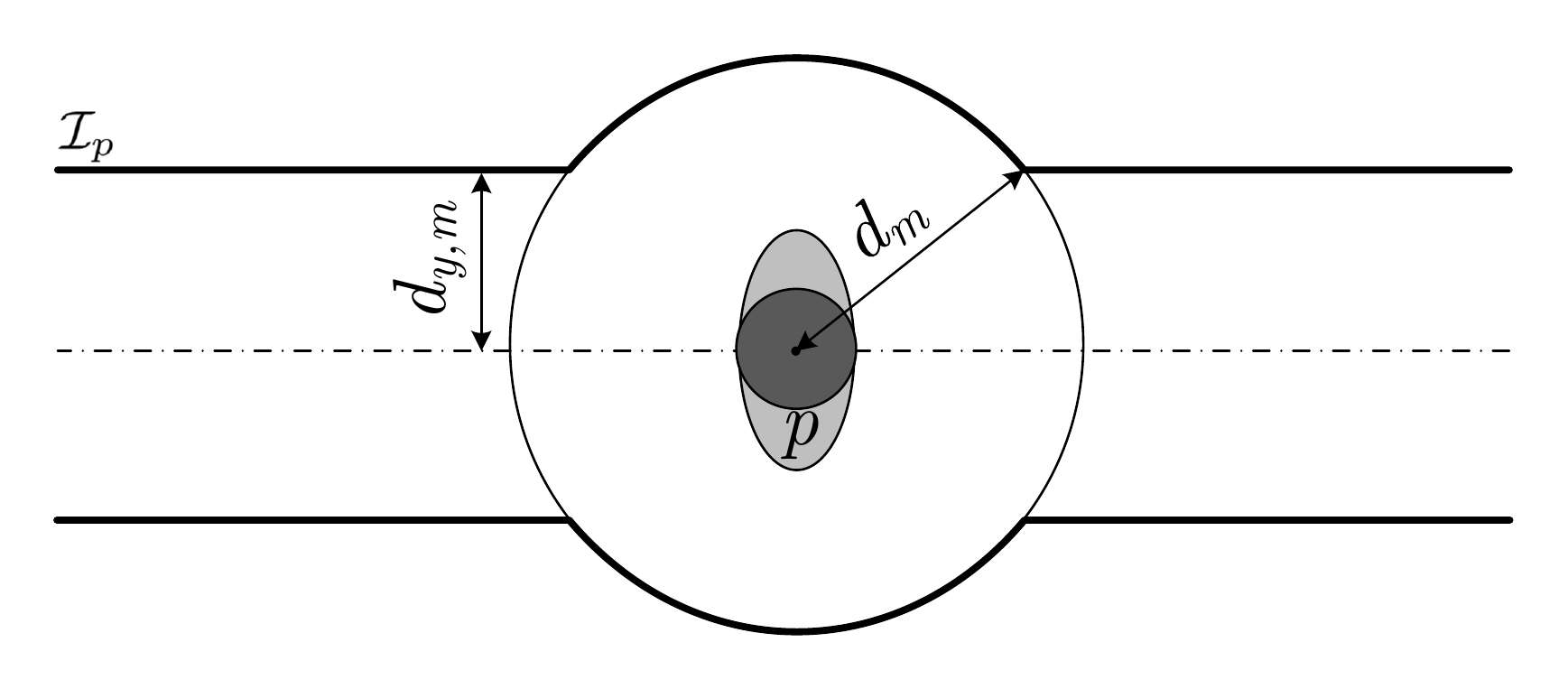}
  \caption{Sketch of the region $\intGrI_p$ around pedestrian $p$. We consider the dynamics of $p$ as 
    potentially being influenced by another pedestrian, say $q$, if, at any time, $q$ entered in  $\intGrI_p$.
    Conversely, if, for a time interval $\tau$  no longer than $\tau_m$,  $q$ entered in  $\intGrI_p$ or, likewise, $p$  entered in $\intGrI_q$, we consider the pair 
    $(p,q)$ as non-interacting (even if the two pedestrians appear in the same frames -- cf. Fig.~\ref{fig:sketch-graph-depth-maps}). 
    By removing the edges in $G$ (cf. $G$ construction in Fig.~\ref{fig:sketch-Graph_creation})
    associated to such non-interacting pedestrians we obtain $\tilde{G}$, in which only potentially interacting pedestrians are connected by edges
    (cf. Sect.~\ref{sect:classification},  
    Tab.~\ref{tab:grap-symbols} and Fig.~\ref{fig:sketch-Graph_process}(B)).
    The region $\intGrI_p$ is parametrized by the lengths $d_m$ and $d_{y,m}$, respectively the minimum length for 
    interaction and the minimum transversal length for interaction -- cf. Tab.~\ref{tab:parameters}). 
  }
       \label{fig:Private_space}
\end{figure}

\begin{figure}[!ht]
\centering
 \includegraphics[width=.45\textwidth, trim={.5cm .5cm .5cm .5cm},clip]{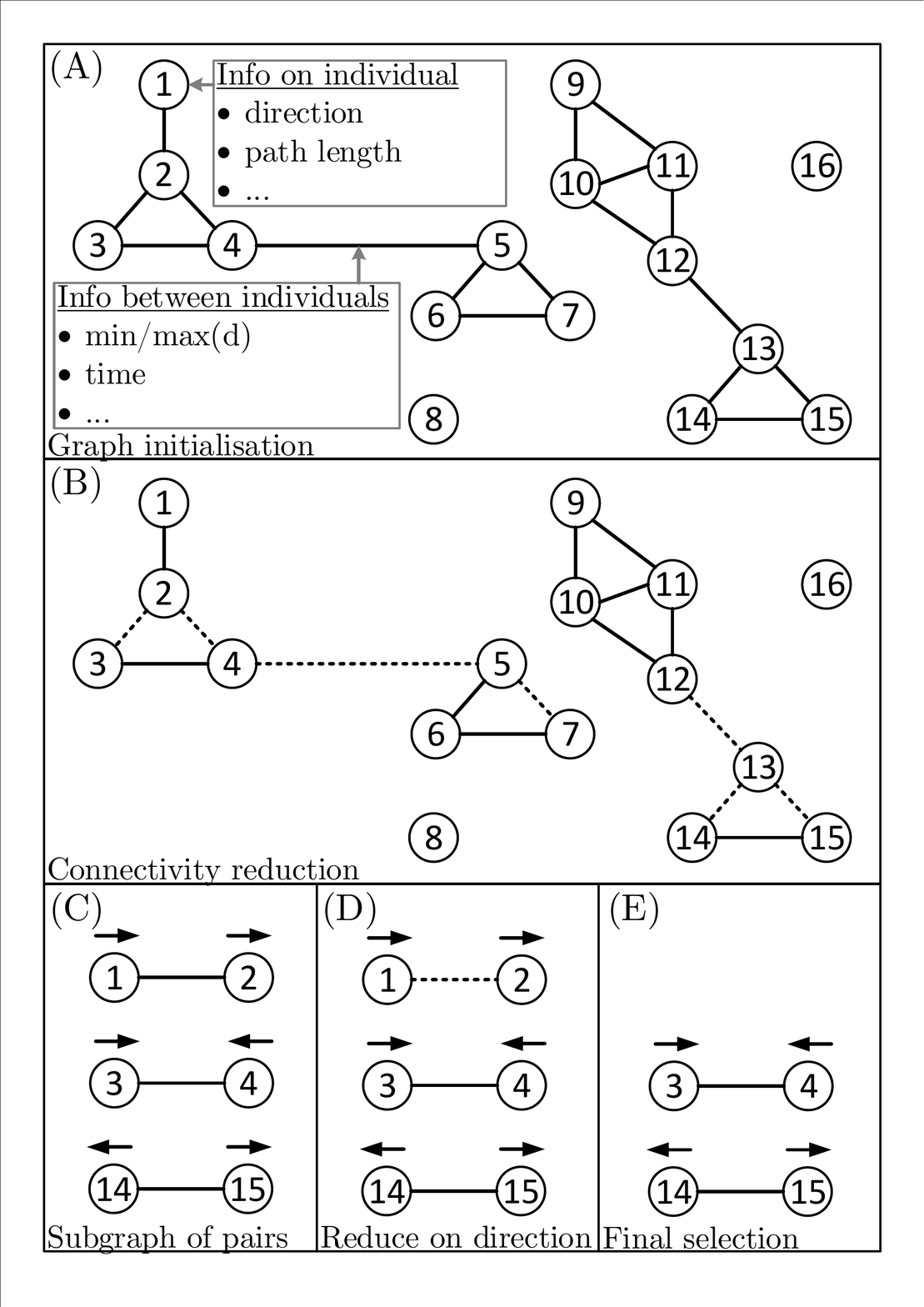}
 \caption{Occurrences of mutual avoidance of two pedestrians walking
   in opposite directions are selected automatically using the graph
   representation. In this figure we summarize the whole process that leads from $G$ (A) to the sub-grap $\tilde{G}_{2,a}$ (E).
   First, the graph representation in (A)   is ``sparisified'' removing edge should two 
   pedestrians be not interacting (according to the condition in Fig.~\ref{fig:Private_space}) yielding the sub-graph $\tilde{G}$ in (B).
   We then isolate the sub-graphs of $\tilde{G}$ constituted only of dyads
   (connected components with two nodes) are isolated (C) to further
   retain only the cases in which the walking direction of pedestrians
   are opposite (D, E). In this last step we further filter to retain
   pairs of pedestrians whose interaction time satisfies $\tau >
   \tau_M$.}  
\label{fig:sketch-Graph_process}
\end{figure}

\subsection{Graph-based representation}\label{sect:representation}
The graph-based representation technique,
described in the previous paragraphs and below, significantly improves what previously proposed by
us in~\cite{corbetta-colldyn17} allowing richer and more parametric
scenario classification (the improvement occurs through node
annotations and edge weighting). 
Because of the bijective correspondences between pedestrians and trajectories, and between trajectories and graph nodes, in the following, 
we will refer interchangeably as pedestrians/trajectories/nodes, which we will identify with the generic symbols $p$ and/or $q$.

We build $G$ as follows: we scan in chronological order the set of experimental trajectories (Sect.~\ref{sect:campaign}), and add a node $p$ to $G$ when a new trajectory is found. 
We further annotate each node with scalar observables of the
trajectory. These are: average walking velocity, trajectory
length, ultimate direction, starting and ending positions.

As we scan the trajectory data we also introduce edges between
nodes. In particular, if pedestrians $p$ and $q$ appear
simultaneously in one or more recorded frames, we add an edge $e = (p,q)$
between the associated nodes.

We vector-weight the edge $e$ with scalars
quantifying the pairwise dynamics of $p$ and $q$. The weight $\vec{w}(e)$ reads
\begin{equation}
\vec{w}(e) = (\min(d),\max(d),\tau),
\end{equation}
where 
\begin{itemize}
\item $d$ is the distance between $p$ and $q$
  (cf. Fig.~\ref{fig:sketch-Trajectories}), of which we retain the
  minimum ($\min$) and the maximum ($\max$) observed values;
\item $\tau$ is the joint recording time, i.e. the duration $p$ and
  $q$ are both present in our recording window.
\end{itemize}
We report the graph construction algorithm in
Fig.~\ref{fig:sketch-Graph_creation}.

We stress that the procedure to construct the graph is efficient in the sense that it is linear-in-time with the amount frames measured: just one pass of $\Gamma$ is necessary to construct $G$.

\begin{figure}[!ht]
 \centering
\includegraphics[width=.225\textwidth]{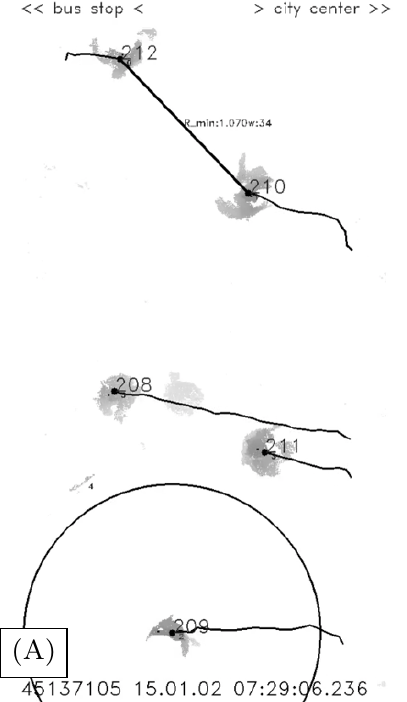}
\includegraphics[width=.225\textwidth]{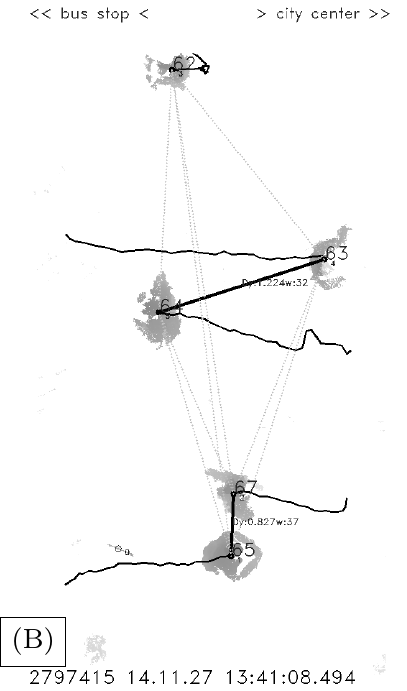}
 \caption{Examples of depth maps collected in our experimental campaign (cf. Fig.~\ref{fig:sketch-Eindhoven_Setup} and Sect.~\ref{sect:campaign}). 
   Individual trajectories have been superimposed to the depth-maps in post-processing. 
   Moreover, the panels include realizations of scenarios of interest.
   Panel (A) contains one realization of pairwise avoidance (pedestrians $(210,212)$, joint by a thick black line) 
   and one undisturbed pedestrian ($209$ - we report the circular
   region of radius $d_m$ around them that remains not visited by others). 
   Panel (B) contains two realizations of pairwise avoidance  (pedestrians $(63,64)$ and $(65,67)$).
   Note that we can have multiple pairwise avoidance realizations in the same frame, modulo there is no interplay between them.
   The graph representation flexibly allows to define these scenarios and efficiently recover them in the measurements.
 }
   \label{fig:sketch-graph-depth-maps}
\end{figure}

\begin{table}
\caption{Parameters employed in the construction of the graph-based representation (Sect.~\ref{sect:graph}), in the undisturbed dynamics model (Sect.~\ref{sect:diluted}) and in its extension to include pairwise avoidance (Sect.~\ref{sec:interactions}). The pairwise avoidance model extends the undisturbed dynamics model and preserve all its parameters but the percentage of runners. This percentage is reduced  following the observations.\label{tab:parameters}}
\begin{ruledtabular}
\begin{tabular}{c l l }
&  & Graph-based representation \\ \hline
$d_m$ & $2.4\,$m & min distance for interaction\\
$d_{y,m}$ & $0.8\,$m & min transversal distance \\
 & &  \quad for interaction\\
$\tau_m$ & $5\,$frames = $1/3\,$s & min time for interaction\\
$\tau_M$ & $20\,$frames = $4/3\,$s & min interaction time \\
& & \quad retained (Sect.~\ref{sect:datasets})\\
\hline
& & Undisturbed dynamics \\
\hline
$u_{p,w}$ & $1.29\,$ms$^{-1}$ & walkers  \\
$u_{p,r}$ & $2.70\,$ms$^{-1}$ &  runners \\
$\sigma_x$ & $0.25\,$ ms$^{-3/2}$ &  \\
$\sigma_y$ & $0.25\,$ ms$^{-3/2}$ &  \\
$\alpha_w$ & $0.037\,$m$^{-2}$s  & walkers \\
$\alpha_r$ & $0.0015\,$m$^{-2}$s &  runners \\
$\beta$ & $1.765\,$ m$^{-2}$s& \\
$\nu$ & $0.297\,$s$^{-1}$ & \\
$y_p$ & $0.0\,$m & \\
runners $\%$   & $4.02$\% & \\
\hline
& & Pairwise avoidance \\
\hline
$\theta_1$ & $20\,^{\circ}$ &  \\
$\theta_2$ & $90\,^{\circ}$ &  \\
$\mu$ & $1.0\,$s$^{-1}$ & \\
$R$ & $2.4\,$m &  \\
$r$ & $0.6\,$m &  \\
$A$ & $1.5\,$ms$^{-2}$ &  \\
$B$ & $0.7\,$ms$^{-2}$ &  \\
runners $\%$   & $0.2\,$\% & \\
\end{tabular}
\end{ruledtabular}
\end{table}

\subsection{Flow classification}~\label{sect:classification}
The representation via $G$ enables us to formally define virtual experiments 
and efficiently classify scenarios, both by
exploiting the graph topology. In this sub-section we introduce the general approach  starting from 
the specific cases of interest for Sect.~\ref{sect:diluted} and~\ref{sec:interactions}.

In Sect.~\ref{sect:diluted} we aim at
analyzing the dynamics of undisturbed pedestrians. Singleton nodes in
$G$ (i.e. edge bereft) provide a first, yet incomplete,  collection of
these pedestrians. In fact, no other pedestrian potentially
perturbing their dynamics was observed while they were crossing our facility 
(the singleton condition would otherwise be violated).

Singleton nodes of $G$, however, identify just a subset of pedestrians walking
undisturbed. In fact, it is reasonable to assume that all individuals
remaining sufficiently far from their first neighbor walked undisturbed as well.
To correctly classify these cases -- heuristically speaking -- we remove edges from
the nodes that are ``sufficiently far apart from their neighbors'' and reduce them to singletons.

More formally, we consider a
reference pedestrian $p$ as potentially influenced by a pedestrian
$q$, if $q$ enters in $p$'s ``neighborhood'' $\intGrI_p$ (to be
geometrically defined below) for at least one frame.

We define the region $\intGrI_p$ (see Fig.~\ref{fig:Private_space})
considering two criteria:
\begin{itemize}
\item pedestrians walking at short distances (say smaller than a given
  threshold $d_m$) most likely play an influence on the respective
  dynamics, therefore
\begin{equation}\label{eq:metric-infl1}
\min(d) < d_m \Rightarrow q \in \intGrI_p;
\end{equation}
\item pedestrian interactions are anisotropic privileging the motion and sight directions over the transversal directions~\cite{helbing2013pedestrian,cristiani2014BOOK}. Therefore, let $d_{y,m}$ a given threshold for the transversal distance, we set
\begin{equation}\label{eq:metric-infl2}
\min(d_y) < d_{y,m} \Rightarrow q \in \intGrI_p.
\end{equation}
(cf. parameters in Tab.~\ref{tab:parameters}).
\end{itemize}
We stress that determining if $q \in \intGrI_p$ or $p \in
\intGrI_q$ consists just of a single check on the vector-weight $\vec{w}(e)$ of the edge $e = (p,q)$.

In case two pedestrians $p$ and $q$, connected by an edge $e$, exerted
no influence on the motion of  each  other (according to the metric criteria
in Eq.~\eqref{eq:metric-infl1}-\eqref{eq:metric-infl2}), we remove the
edge $e$. This operation returns a ``sparsified'' sub-graph $\tilde{G} \subset G$, likely with 
an increased number of  singletons.
Let us call $\tilde{G}_1$ the sub-graph of singletons of $\tilde{G}$.
$\tilde{G}_1$ identifies all the realizations of undisturbed flows, i.e.  
 the experimental data for our analysis in Sect.~\ref{sect:diluted}
(further technical constraints on the dataset are described in
Sect.~\ref{sect:datasets}).

As we expect all
singleton nodes to be associated to a similar dynamics (undisturbed pedestrians), 
we expect connected components with 
similar edge topology, weights and annotation, to exhibit similar
dynamics, and thus to be realizations of the same scenario. 
In this sense, we formally define  virtual experiments by specifying an edge topology and ranges for weights and annotations.
This selects a sub-graph  $\tilde{G}_s \subset \tilde{G}$, 
and all the connected components of $\tilde{G}_s$ are associated to the realization of the scenario.

We exploit this concept to retain data about avoidance
dynamics of pairs (Sect.~\ref{sec:interactions}).
We find pairs of pedestrians in avoidance among the dyads (connected
components of two nodes) in $\tilde{G}$. Specifically, we retain only
those dyads in which
\begin{inparaenum}
\item the walking directions of the two pedestrians are opposite;
\item the two pedestrian initially faced each other.
\end{inparaenum}
We call $\tilde{G}_{2,a}$ the sub-graph of such dyads, to be used in the analyses in Sect.~\ref{sec:interactions}.
(As in the case of $\tilde{G}_1$, also here we consider further technical
constraints, discussed in Sect.~\ref{sect:datasets}).

In Fig.~\ref{fig:sketch-Graph_process}, we report a schematic description of
the selection of avoidance pairs, while some examples of real
data selected via the procedure reported here are in 
Fig.~\ref{fig:sketch-graph-depth-maps}.

We finally stress that we use pairwise metric properties
(i.e. $\vec{w}(e)$) as discriminant of the occurrence of an
interaction. It has been recently
recognized that pedestrians interactions, especially at high
densities, can be determined by factors beyond the sole
metric~\cite{Mou-pnas} as it happens, e.g., for social
animals~\cite{ballerini2008interaction}. As here we restrict to
free-flow conditions and to one-to-one interactions, analyses based on
metric arguments appear sufficient.


\begin{figure*}[t]
 \centering
\includegraphics[width=0.31\textwidth]{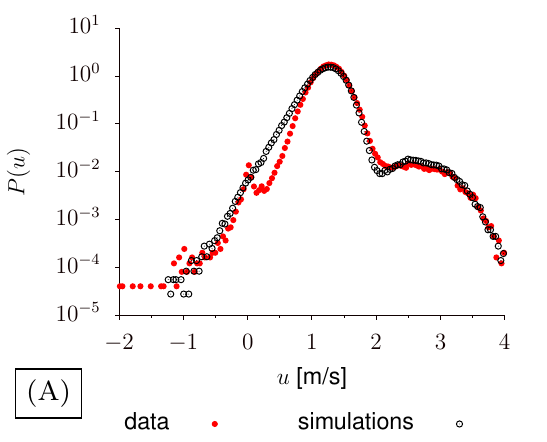}
\includegraphics[width=0.31\textwidth]{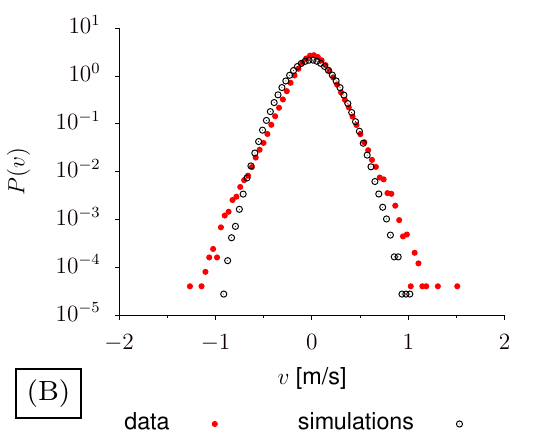}
\includegraphics[width=0.31\textwidth]{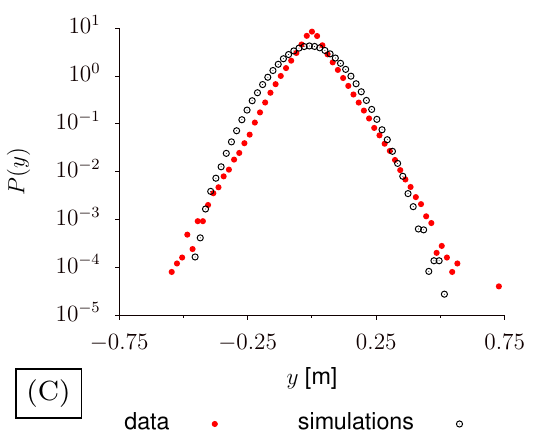}
  \caption{Probability distribution functions of walking velocity and
    positions for undisturbed individuals: comparison between
    measurements (solid red dots) and simulations of
    Eq.~\eqref{subeq1:SPM}-\eqref{subeq4:SPM} (black empty dots,
    simulation parameters in Tab.~\ref{tab:parameters}). The panels
    contain respectively (A) longitudinal velocities ($u$), (B) transversal
    velocities ($v$), (C) transversal positions ($y$, assuming $y_p =0$). 
    Pedestrians walk most
    frequently at around $1.29\,$m/s (cf. (A)). Besides, we observe a
    tiny fraction of running pedestrians, about $4\,\%$,
    contributing to the hump at above $2\,$m/s and pedestrians turning
    back, providing negative velocities contributions. Transversal
    fluctuations in velocity (B) appear well-approximated by a Gaussian
    distribution, while transversal positions exhibit small deviations from 
    a Gaussian behavior (C).
Our model (Eq.~\eqref{subeq1:SPM}, ~\eqref{subeq3:SPM}) reproduces quantitatively the
complete longitudinal velocity statistics inclusive of the running
hump as well as the inversion events. The transversal dynamics is also
well approximated as a stochastic damped harmonic oscillator
(Eq.~\eqref{subeq2:SPM}, ~\eqref{subeq4:SPM}).} 
       \label{fig:PDF_Velocity_comparison}
\end{figure*}


\subsection{Datasets}\label{sect:datasets}
In this technical section we discuss the restrictions and transformations
further applied to the trajectory sets selected by $\tilde{G}_1$ and
$\tilde{G}_{2,a}$ to yield the datasets employed in the next sections. The restrictions specified below identify further
sub-graphs within $\tilde{G}_1$ and $\tilde{G}_{2,a}$.  We  however refrain from introducing new symbols and maintain,
with an abuse of notation, these identifiers for the sub-graphs.

\noindent  \textbf {Dataset for diluted flow analysis} (Sect.~\ref{sect:diluted} - constraints on $\tilde{G}_1$):
  \begin{itemize}
  \item \textit{Restriction to straight intended paths and quasi-rectilinear trajectories.}  
    We aim at
    analyzing the fluctuations of the undisturbed motion when occurring
    around  intended paths that are straight. 
    In these conditions, we expect to observe trajectories that are quasi-rectilinear.
    From energy minimization arguments
    (cf.~\cite{arechavaleta2008optimality}), we expect intended paths to be straight 
    when it comes to reach targets in
    obstacle-free environments. 
    In our dataset, we could however also observe largely
    erratic trajectories formed e.g. of circular sections
    or of parabolic arcs. As these trajectories are out of our modeling purpose, 
    we discard them after identify them through the procedure in Appendix~\ref{app:coord-single}.
  \item \textit{Coordinate system.} Limiting our scope to 
    quasi-rectilinear trajectories, we rotate them for convenience such that,
    in a $(x,y)$ reference system, $x$ is the longitudinal walking
    direction and $y$ is the direction of the transversal
    fluctuations (quasi-rectilinear trajectories have generally different inclinations depending on their starting position, cf. Fig.~\ref{fig:quiver_binned_statistic_2d}). 
    The details of the rotation procedure are in
    Appendix~\ref{app:coord-single}.
  \end{itemize}

\noindent \textbf{Dataset for pairwise avoidance analysis} (Sect.~\ref{sec:interactions} - constraints on $\tilde{G}_{2,a}$):
  \begin{itemize}
  \item \textit{Time thresholding of pairwise dynamics.}  Due to the
  finiteness of our observation window, the joint observation time for
  a pairwise dynamics can be limited to few frames. For instance,
  this occurs when a pedestrian of the pair is about to leave as the
  second enters the domain. To exclude these cases from our dataset we
  impose a lower bound, $\tau_M$, on the joint recorded time, i.e. we
  require $\tau > \tau_M$. We choose $\tau_M$ to be comparable with
  the crossing time of the observation window of an undisturbed
  pedestrian. Notably this restriction guarantees that the point in
  space at which the two pedestrians of the pair are closest is
  roughly in the middle of the observation window.
  
\end{itemize}

\section{Undisturbed motion}\label{sect:diluted}
In this section we model the dynamics of pedestrians walking undisturbed (also referred to as \textit{free flow}), keeping as a quantitative reference the measurements collected from the setup in Fig.~\ref{fig:sketch-Eindhoven_Setup} (cf. Sect.~\ref{sect:campaign}) and selected through $\tilde{G}_1$ (cf. Sect.~\ref{sect:classification} -- ~\ref{sect:datasets}).
The free flow motion is a limit condition for the dynamics, as it involves pedestrian densities
at its lowest levels. We consider it as a reference condition for which we interpret pairwise interactions (analyzed in Sect.~\ref{sec:interactions}) as perturbations. We proceed to deduce a model for undisturbed conditions and we then compare it with our measurements in terms of probability distribution functions.

Individuals crossing a large corridor, typically move following
straight \textit{intended} paths along quasi-rectilinear trajectories (cf. Sect.~\ref{sect:datasets}).
Besides this variability in the individual intended paths, each pedestrian performs small and high-frequency random fluctuations (about $1\,$Hz due to  walking physiology). 
Moreover, as observed in~\cite{corbetta2016fluctuations},
rare large fluctuations in the motion occur too. Such rare large deviations
include, but does not limit to,  trajectory
inversions. In~\cite{corbetta2016fluctuations}, these two apparently
independent fluctuating phenomena have been treated as realization of
a unique Langevin stochastic dynamics with an 
bi-stable longitudinal velocity potential. In other words, the
dynamics was treated in terms of a longitudinal velocity, $u$,
exhibiting small fluctuations around a stable state, $u = u_p$, plus
occasional velocity inversion events, $u \rightarrow -u$ (for which the
dynamics stabilizes on $u = -u_p$).

Here we provide a twofold extension of the Langevin model
in~\cite{corbetta2016fluctuations} for the wider, longer and less
constrained walking area considered
(cf. Fig.~\ref{fig:sketch-Eindhoven_Setup}). For the sake of
completeness, we first present the extended model and then we discuss
it in view of our previous work~\cite{corbetta2016fluctuations}.

For convenience, we adopt a coordinate system $(x,y)$, where $x$ is a
longitudinal coordinate along the walking direction, i.e. along the intended path, which we consider as a straight line parallel to the longitudinal direction of the corridor, and parametrized by the variable $y_p$. The variable $y$ accounts
for transversal position, so that $y-y_p$ identifies the fluctuation around the intended path.
 We model the longitudinal and
transversal dynamics as uncorrelated Langevin motions satisfying
\begin{align}
\frac{\mathrm{d}x}{\mathrm{d}t} 				&= u(t) \label{subeq1:SPM}\\
\frac{\mathrm{d}y}{\mathrm{d}t} 				&= v(t) \label{subeq2:SPM}\\
\frac{\mathrm{d}u}{\mathrm{d}t} 				&= f(u)  + \sigma_x\dot{W}_x  \label{subeq3:SPM}\\
\frac{\mathrm{d}v}{\mathrm{d}t} 				&= - 2 \nu v(t) + \beta g(y)  + \sigma_y \dot{W}_y \label{subeq4:SPM}
\end{align}
where $u$ and $v$ are, respectively, the transversal and longitudinal
velocity components, $\nu$ and $\beta$ are positive model
parameters, $\dot{W}_x$ and $\dot{W}_y$ are independent,
$\delta$-correlated in time Gaussian noise scaled by the
positive coefficients $\sigma_x,\sigma_y$ (we assume, consistently with~\cite{corbetta2016fluctuations}, $\sigma_x = \sigma_y$, c.f. Tab.~\ref{tab:parameters}.  Our choice for the noise is common and
made for simplicity, yet it is not mandatory. See
e.g.~\cite{Lutz}). The features of the dynamics are finally
incorporated in the two functions $f(u)$ and $g(y)$.

As in~\cite{corbetta2016fluctuations}, we choose $f(u)$ as possibly
the simplest smooth model for a bi-stable dynamics, i.e. as the
gradient of a double well (velocity) potential. In formulas it reads
\begin{equation}\label{eq:long-pot}
f(u) = - 4 \alpha_i u (u^2-u_{p,i}^2) = - \partial_u \alpha_i (u^2-u_{p,i}^2)^2,
\end{equation}
with $\pm u_{p,i}$ being the expected stable velocities, and
$\alpha_i$ as the modulating factor of the force. As a first extension
of the model in~\cite{corbetta2016fluctuations}, we introduce the
subscript, $i$, to enable multiple populations all behaving identically
except for the stable velocity value. This allows one to distinguish
e.g. people walking at usual speed and runners.

The function $g(y)$ models the restoring impulse towards the intended
path.  In formulas, it reads
\begin{equation}
g(y) = - 2 (y-y_{p})  = - \partial_y (y-y_{p})^2.
\end{equation}
This marks a second, yet fundamental, extension to the model
in~\cite{corbetta2016fluctuations}.  In fact, a wide corridor enables
a continuous choice of straight intended paths that remain unchanged
during the motion (in formulas, $\dot y_p = 0$). In turn, in
Sect.~\ref{sec:interactions}, we describe interactions considering a dynamics also for the variable $y_p$.

In Fig.~\ref{fig:PDF_Velocity_comparison}, we compare the measured and
modeled pedestrian motion in terms of probability distribution
functions of longitudinal and transversal velocity and transversal fluctuations with respect to the intended path. 
The figures
include data from $N=47\,122$ trajectories of average time length
$2\,$s (i.e. $31$ frames). Approximately $34\,\%$ of the trajectories
are from undisturbed pedestrians walking towards the bus terminal 
(with the rest are from undisturbed pedestrians walking towards the city center). 
The comparison is performed with $47\,122$  trajectories simulated via~\eqref{subeq1:SPM}-\eqref{subeq4:SPM}, 
and calibrating the parameters as in~\cite{corbetta2016fluctuations}
(values reported in  Tab.~\ref{tab:parameters}).

In the longitudinal velocity ($u$) probability
distribution (Fig.~\ref{fig:PDF_Velocity_comparison}(A)) we
observe different regimes: most likely people walk with speed fluctuating around
 $1.29\,$m/s. Moreover, about $4\,\%$ of the
pedestrians run across the walkway: this results in the hump at the
right hand side of the distribution. Finally, rare events such as
turning backs trajectories and stopping are present which provide,
respectively, contributions in the left tail and at around
$0\,$m/s. By adopting the measured ratio of walkers and runners, 
simulations quantitatively reproduce the observed velocity distribution.
We observe Gaussian transversal fluctuations of the velocity ($v$) 
that Eq.~\eqref{subeq2:SPM} and~\eqref{subeq4:SPM} capture. Slight deviations from the
 predicted Gaussian fluctuations in transversal position ($y$) are instead observed.

\section{Pairwise avoidance}\label{sec:interactions}
In this section we model  the dynamics of the pairwise avoidance and of the related changes in intended path (cf. conceptual sketch in Fig.~\ref{fig:sketch-Trajectories} and measured cases in Fig.~\ref{fig:sketch-graph-depth-maps}). We consider these dynamics in the simplest condition involving exclusively two pedestrians walking in opposite direction and avoiding each other while remaining sufficiently far (i.e. not influenced) from any other individuals. In this sense, we deal with avoidance in diluted conditions.
We compare with our measurements selected through the (connected components of the) sub-graph $\tilde{G}_{2,a}$ (cf. Sect.~\ref{sect:classification} --\ref{sect:datasets}.  Each of the two nodes of the dyads in   $\tilde{G}_{2,a}$ corresponds to one of the two pedestrians involved). This scenario represents for us the first building block to treat quantitatively interaction dynamics on top of the undisturbed motion (Sect.~\ref{sect:diluted}). In the following we first describe our model, then we compare it with the measurements in terms of probability distributions.

To model the pairwise avoidance we consider two individuals, each modeled following Eq.~\eqref{subeq1:SPM}-\eqref{subeq4:SPM} plus coupling terms affecting both individual positions $(x,y)$ and individual planned paths $y_p$. In other words, the individual state is now described by the triplet $(x,y,y_p)$ (and derivatives), with $y_p$ entering in the dynamics as a variable and not as a constant parameter. We model the coupling terms as social forces~\cite{helbing1995PRE,parisi2009modification}, acting on the whole triple $(x,y,y_p)$.

\begin{figure}[!ht]
 \centering
  \includegraphics[width=.35\textwidth, trim={.5cm .5cm .5cm .5cm},clip]{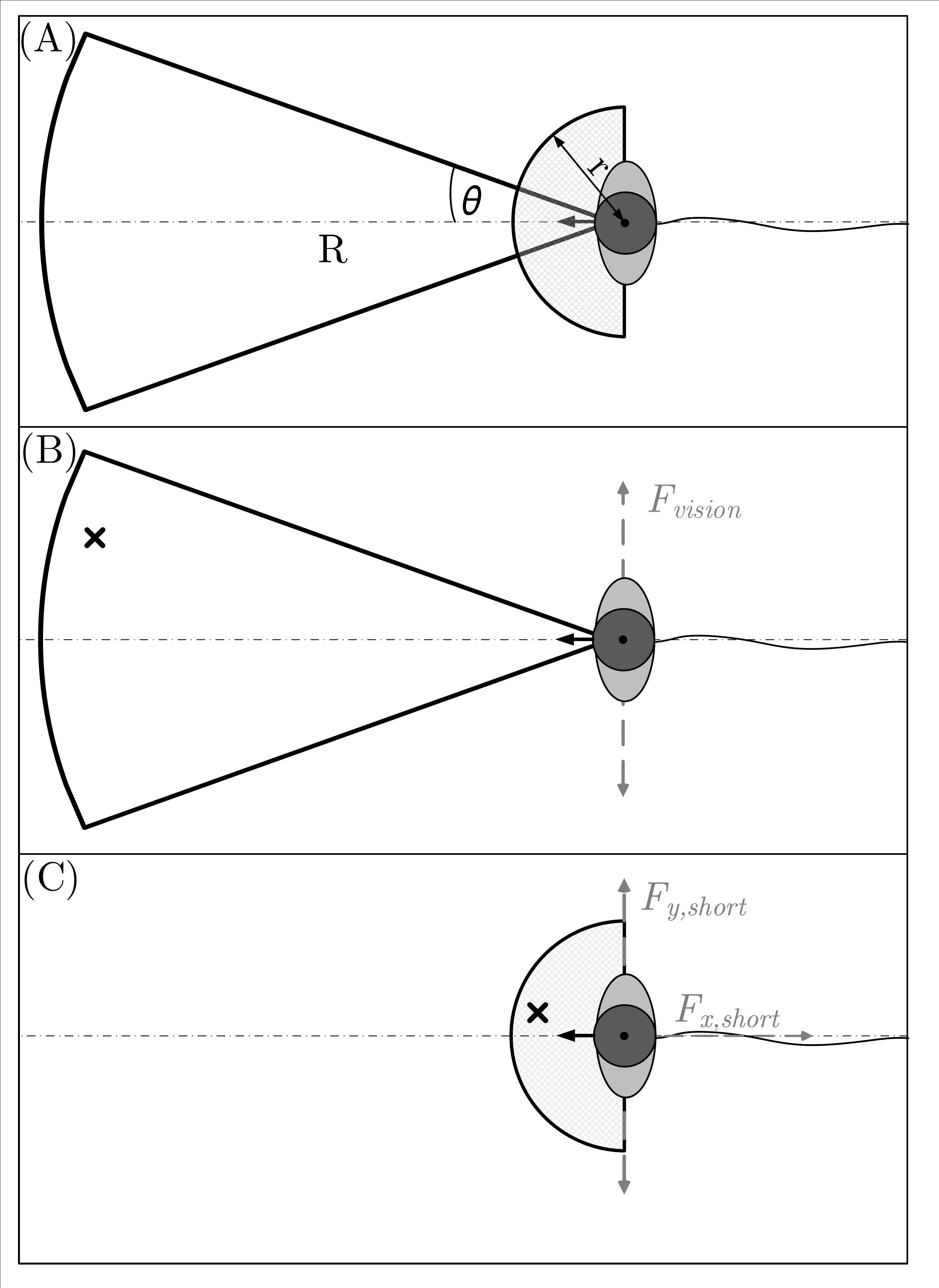}
  \caption{Schematics of the interaction forces considered. Sketch (A): The
    intended path of a pedestrian is modified on a twofold basis: by
    long-range, sight-based (thus anisotropic) forces
    (cf. Eq.~\eqref{eq:Fvisiony}) and by short-range contact avoidance
    forces (cf. Eq.~\eqref{eq:Fshortx}). Long range forces are bounded
    within a circular sector of radius $R$ and angular semi-amplitude
    $\theta$ located in front of the individual (i.e. aligned with the
    intended path).  Short-range forces are frontal and
    bounded within a circular region of radius $r$.  Sketch (B): Long-range
    sight-based interactions, e.g. with the pedestrian marked with a
    ``$\times$'', yield forcing $F_{vision}$ in orthogonal direction
    with respect to the intended path.  Sketch (C): Short-rage forces provide
    isotropic $F_{short,x} = F_{short,y}$, although frontal only,
    contact avoidance, e.g. of the pedestrian marked with a
    ``$\times$''.} 
\label{fig:sketch-Forces}
\end{figure}

Let $p_1$, $p_2$ be two pedestrians in an avoidance event.
Adopting the point of view of one of the two, say $p_1$, and using the same $(x,y)$ reference system used in
Sect.~\ref{sect:diluted}, we model the dynamics as
\begin{align}
\frac{\mathrm{d}y_{p}}{\mathrm{d}t} 			&= \dot{y_{p}}(t) \label{subeq1:Diff_model_int} \\
\frac{\mathrm{d}\dot{y}_{p}}{\mathrm{d}t} 	&=  F_{vision} - 2 \mu \dot{y}_{p}(t) \label{subeq2:Diff_model_int} \\
\frac{\mathrm{d}x}{\mathrm{d}t} 				&= u(t) \label{subeq3:Diff_model_int}\\
\frac{\mathrm{d}y}{\mathrm{d}t} 				&= v(t) \label{subeq4:Diff_model_int}\\
\frac{\mathrm{d}u}{\mathrm{d}t} 				&= - 4 \alpha_i u (u^2-u_{p,i}^2) + \sigma_x\dot{W}_x - e_x F_{short} \label{subeq5:Diff_model_int}\\
\frac{\mathrm{d}v}{\mathrm{d}t} 				&= - 2 \nu v - 2 \beta (y-y_{p}) + \sigma_y\dot{W}_y - e_y F_{short}  + F_{vision} \label{subeq6:Diff_model_int}
\end{align}
where the unit vector of components $(e_x,e_y)$ is directed from $p_1$ to $p_2$
 and $\mu$ is a positive model parameter. We
superimpose to the undisturbed dynamics in
Eq.~\eqref{subeq1:SPM}-\eqref{subeq4:SPM} two (social) forces: $F_{short}$ and $F_{vision}$, encompassing
respectively for two influencing elements of the interaction
dynamics. $F_{short}$ is a short-ranged contact avoidance
force; it mimics one's immediate and strong collision-avoidance
reaction to individuals in the very vicinity and acts on the velocity
variables $u,v$. $F_{vision}$, in turn, mimics the sight based
avoidance maneuvers having longer and anisotropic range. $F_{vision}$
acts in the transversal direction only, affecting both, and equally,
the transversal velocity $v$ and the intended path $y_p$.  This
modeling choice follows the idea confirmed from measurements
 that avoidance not only yields lateral motion, but also
provides a persistent change of our intended paths.

We model both forces with decaying exponential of the squared distance between pedestrians (as common practice in the pedestrian dynamics community~\cite{helbing1995PRE}). In formulas, they read
\begin{align}
&F_{vision} = -\mathrm{sign}(e_y) A \, \mathrm{exp}(-d^2/R^2) \chi_1(\tilde \theta) \label{eq:Fvisiony}\\
&F_{short} =  B \, \mathrm{exp}(-d^2/r^2)\chi_2(\tilde \theta) \label{eq:Fshortx},
\end{align}
where $A$ and $B$ are positive parameters, $d$ is the (scalar) distance between the two considered pedestrians (cf. Fig.~\ref{fig:sketch-Trajectories}), $r$ and $R$ are scaling factors for the interaction ranges,  $\tilde \theta$ is the angle between the line joining the two pedestrians and the horizontal and  $\chi_j(\tilde\theta) = 1$ for $|\tilde \theta| < \theta_j$ and $0$ otherwise  ($j=1,2$).

\begin{figure}[!ht]
\centering
\includegraphics[width=.45\textwidth, trim={.5cm .5cm .5cm .5cm},clip]{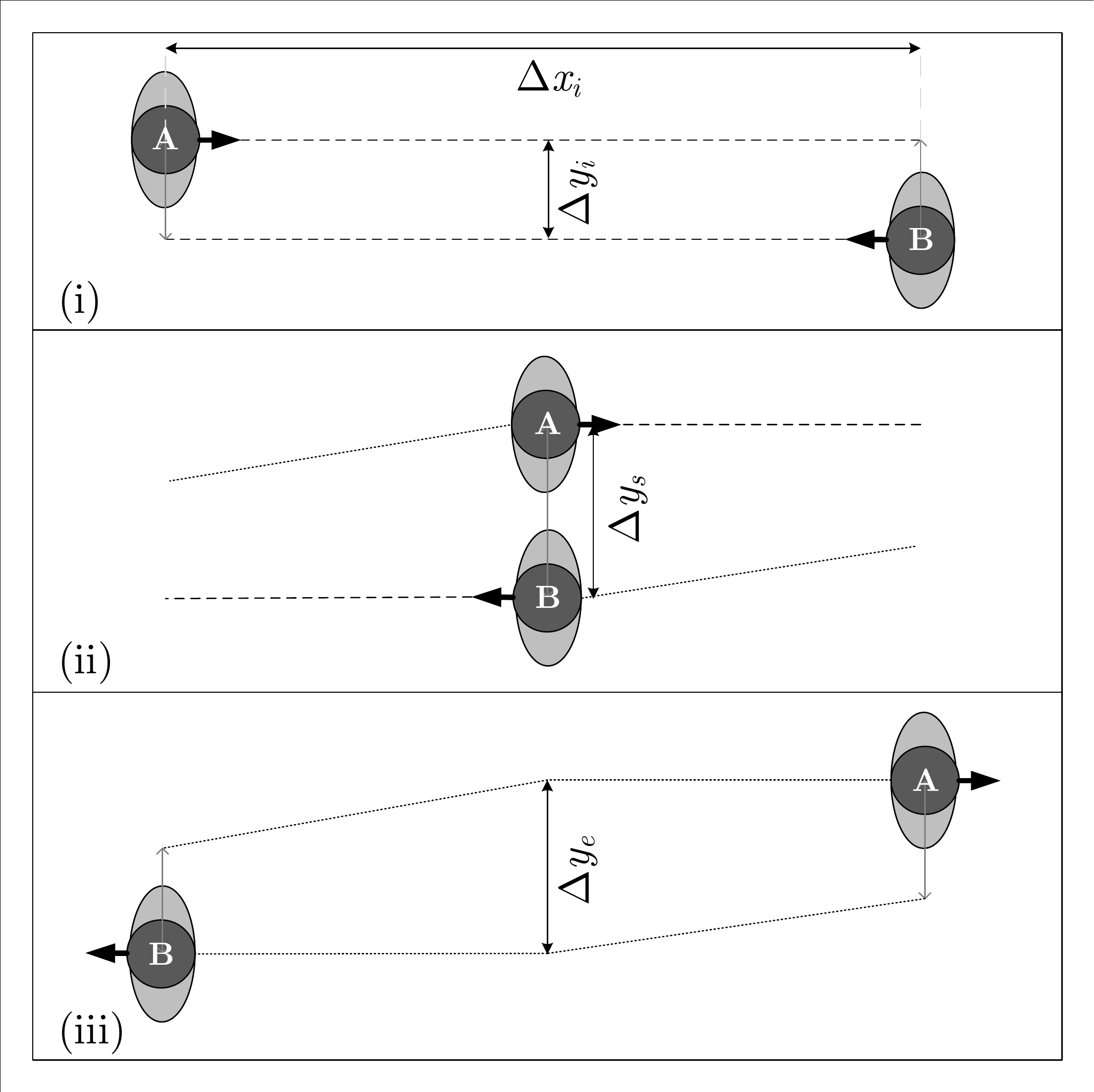}
  \caption{Considered phases of the counter-flowing pairwise dynamics for the scatter plots in Fig.~\ref{fig:comparison-IC} and~\ref{fig:comparison-CF}. 
Sketch (i): Entrance, i.e. the first moment of simultaneous appearance of the pedestrian pair in our observation window.  Sketch (ii): side-by-side walking. it occurs when the two pedestrians have the same longitudinal position.   Sketch (iii): exit, i.e. the last moment in which the two pedestrians appear together.  $|\Delta y_i|$, $|\Delta y_s|$  and $|\Delta y_e|$ indicate the absolute lateral distance between the two pedestrians in phases (i), (ii) and (iii), respectively. }
\label{fig:sketch-config}
\end{figure}

The coupled systems of Langevin equations (Eq.~\eqref{subeq1:Diff_model_int}-~\eqref{subeq6:Diff_model_int}) for $p_1$ and $p_2$ feature, as a whole, a one-dimensional translational symmetry group. Let $(x_1,y_1,y_{p,1})$ and $(x_2,y_2,y_{p,2})$ be the state, respectively, of $p_1$ and $p_2$, the symmetry reads
\begin{equation*}
y_1,\, y_{p,1},\,y_2,\,y_{p,2} \rightarrow y_1 +c,\, y_{p,1} +c,\, y_2 +c,\, y_{p,2} +c 
\end{equation*}
for any real number $c$. In other words, the dynamics is invariant modulo a rigid translation of the transversal position and planned path of both individuals.

\begin{figure*}[t]
\centering
\includegraphics[width=0.45\textwidth]{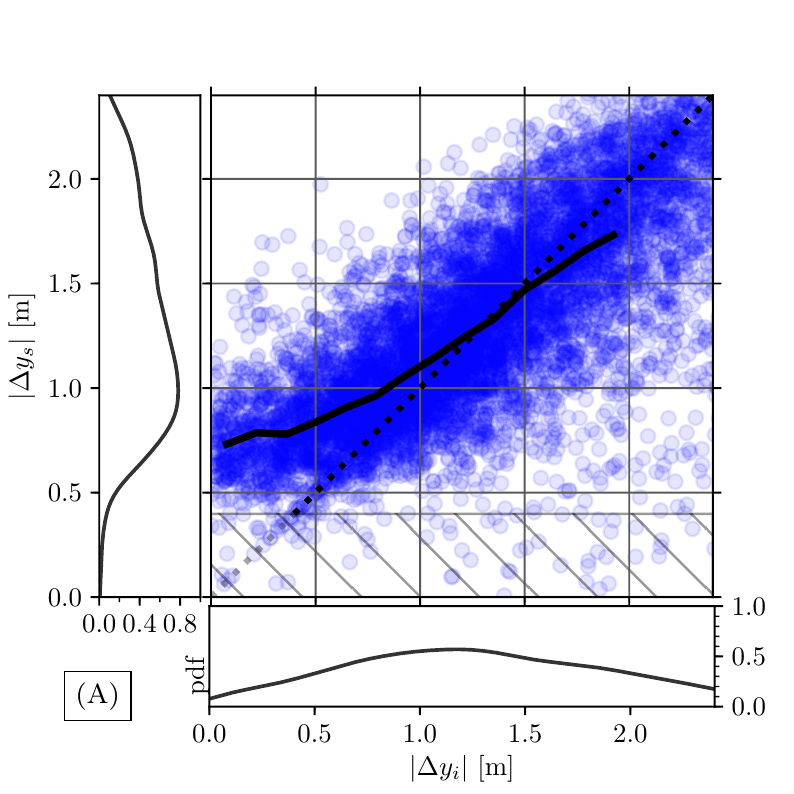}
\includegraphics[width=0.45\textwidth]{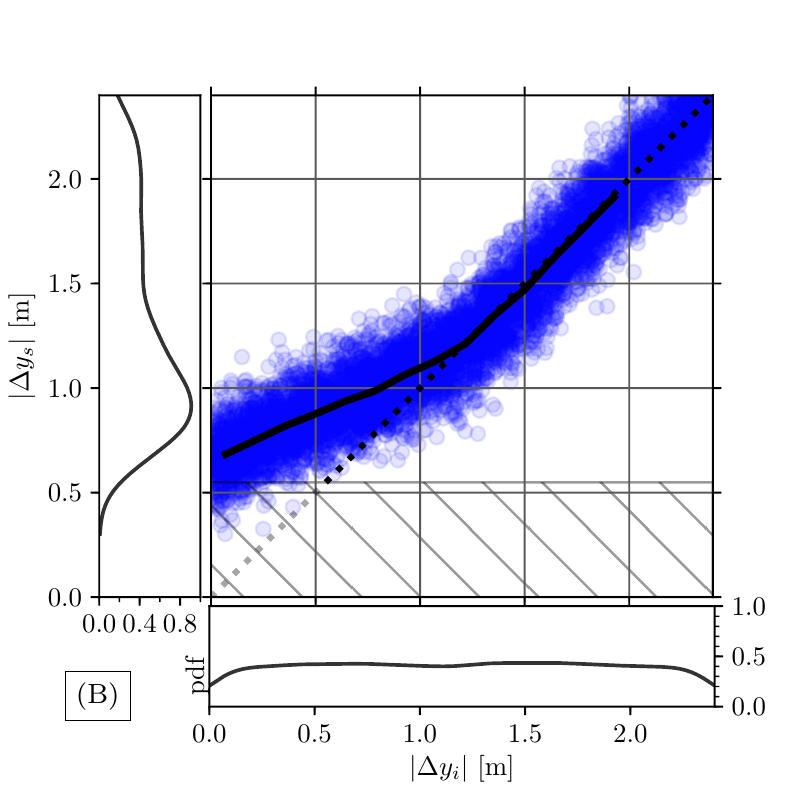}

  \caption{Absolute lateral distance at entrance ($|\Delta y_i|$,
    x-axis, cf. Scenario (i) in Fig.~\ref{fig:sketch-config})
    vs. absolute lateral distance when side-by-side ($|\Delta y_s|$,
    y-axis, cf. Scenario (ii) in Fig.~\ref{fig:sketch-config}). Each
    sample in the scatter plot represents a measured or a simulated
    pair of pedestrians in counterflow (respectively in panels (A) and (B)). The function $e(|\Delta y_s|)$
    (i.e. the ensemble average value of $|\Delta y_s|$ conditioned
    to  $|\Delta y_i|$, cf. Eq.~\eqref{eq:ensavg-trend}) is reported as a solid line. 
    The diagonal $|\Delta y_s| = |\Delta y_i|$ represents
    dynamics for which the lateral distance of a pedestrian pair
    remains unchanged between scenarios (i) and (ii) (in formulas $\frac{d \Delta y}{dt} = 0$).
    In other words, both
    pedestrians walked straight (modulo a constant lateral
    offset). $e(|\Delta y_s|)$ departs from the diagonal for low
    $|\Delta y_i|$ which identifies pedestrians walking towards each
    other thus avoiding each other. As $|\Delta y_i|$ grows,
    interactions/collisions vanish, thus the asymptotic tendency toward
    the diagonal line. A synthetic comparison limited to the
    average trends (Eq.~\eqref{eq:ensemble-approx-yp}) is reported in
    Fig.~\ref{fig:comparison-impactP}(A).  The bottom and left subpanels in (A) and (B) report 
    the (marginal) distributions of $|\Delta y_i|$ and $|\Delta y_s|$.   } 
\label{fig:comparison-IC}
\end{figure*}

\begin{figure*}[t]
 \centering
\includegraphics[width=0.45\textwidth]{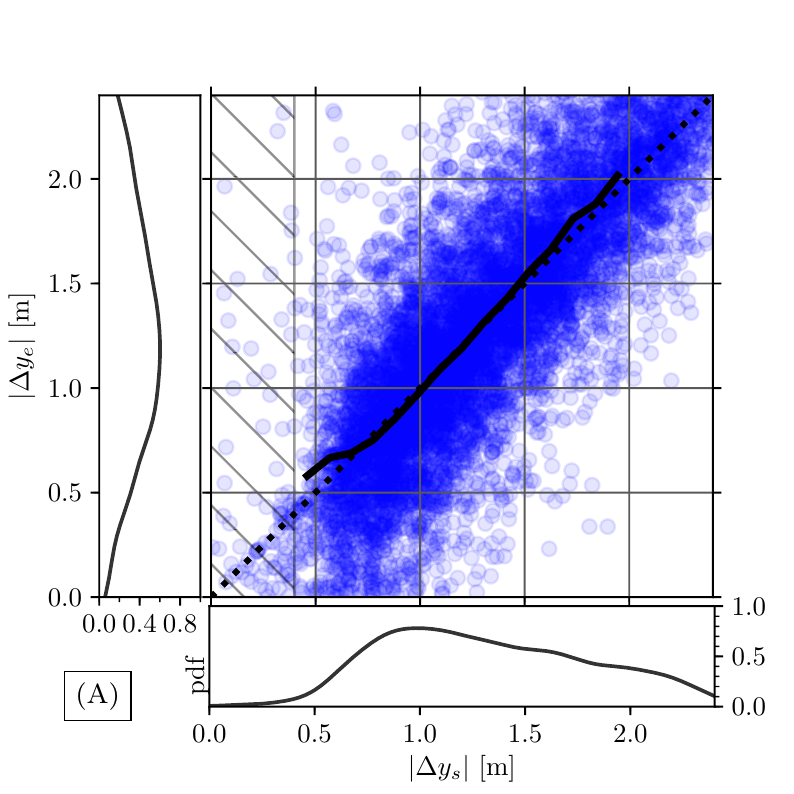}
\includegraphics[width=0.45\textwidth]{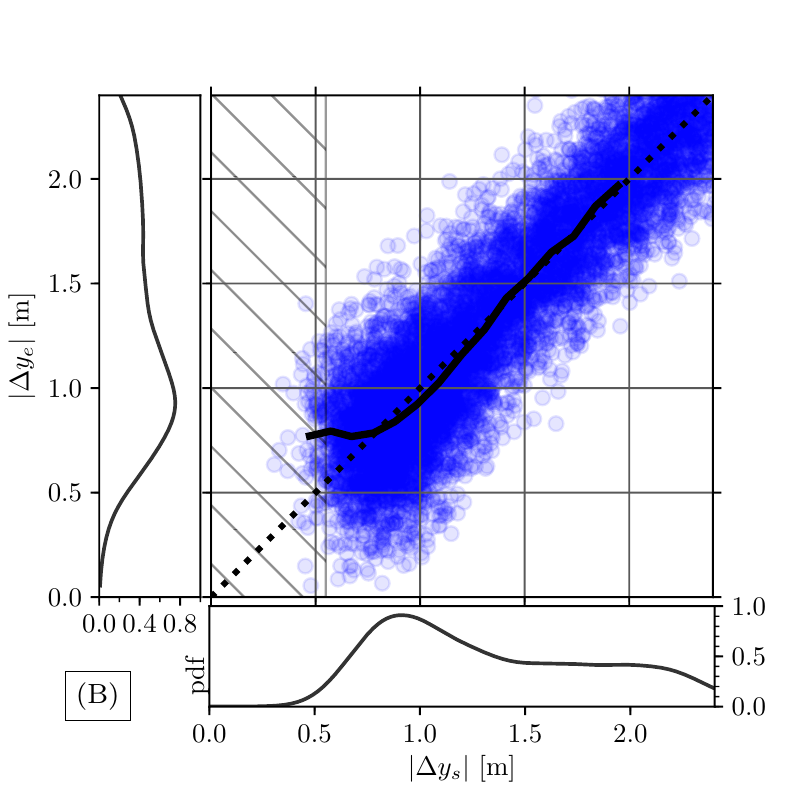}

  \caption{Absolute lateral distance when side-by-side ($|\Delta
    y_s|$, x-axis, cf. Scenario (ii) in Fig.~\ref{fig:sketch-config})
    vs. absolute lateral distance at the exit ($|\Delta y_e|$, y-axis,
    cf. Scenario (iii) in Fig.~\ref{fig:sketch-config}). Each sample in
    the scatter plot represents a measured or a simulated pair of
    pedestrians in counterflow (respectively in panels (A) and (B)).
    Once avoidance is ensured the lateral distance is
    maintained until the exit from our observation window, as the
    scatter samples concentrates around the diagonal $|\Delta y_s| =
    |\Delta y_e|$. A synthetic comparison including only the average
    trends (Eq.~\eqref{eq:ensemble-approx-yp}) is reported in
    Fig.~\ref{fig:comparison-impactP}B. }
       \label{fig:comparison-CF}
\end{figure*}

We compare model and data first in terms of the absolute transversal distance between $p_1$ and $p_2$: $|\Delta y| = |y_1 -
y_2|$ (cf. Fig.~\ref{fig:sketch-config}. Similarly to the undisturbed case discussed in Appendix~\ref{app:coord-single}, pairs of measured trajectories underwent also a rotation procedure to align with our coordinate system. Details are reported in App~\ref{app:coord-pairs}). 
We expect $|\Delta y|$ to well approximate the transversal distance
between the intended paths $|y_{p,1}-y_{p,2}|$ once  considered at the ensemble level, i.e.
\begin{equation}\label{eq:ensemble-approx-yp}
\ensavg{\Delta y} = \ensavg{|y_1 - y_2|}  \approx \ensavg{|y_{p,1}-y_{p,2}|},
\end{equation}
where $\ensavg{\cdot}$ denotes an ensemble average over the measurements (we indicate with $\Eavg_{ens}\left[\cdot \big| E \right]$, the conditioned ensemble average, where  $E$ is the conditioning event). 
We expect, in fact, individual fluctuations with respect to the intended paths be negligible after the (ensemble-)average.

We measure $|\Delta y|$ at three phases of the interaction: 
\begin{itemize}
\item the first appearance of $p_1$ and $p_2$ in our observation window ($|\Delta y_i|$ in Fig.~\ref{fig:sketch-config}(i))
\item the time instant, in the following referred to as $t_s$, when the pair is side-by-side  ($|\Delta y_s|$ in Fig.~\ref{fig:sketch-config}(ii))
\item the last simultaneous appearance of the pair in our observation window ($|\Delta y_e|$ in Fig.~\ref{fig:sketch-config}(iii)).
\end{itemize}
We report our measurements and simulations results in terms of two scatter plots targeting 
two halves of the interaction dynamics: S1, the
avoidance maneuvers until the side-by-side moment (plane $(|\Delta y_i|, |\Delta y_s|)$,
Fig.~\ref{fig:comparison-IC}), and S2, the regime following (plane $(|\Delta y_s|, |\Delta y_e|)$,
Fig.~\ref{fig:comparison-CF}). 

\begin{figure*}[t]
\centering
\includegraphics[width=0.45\textwidth]{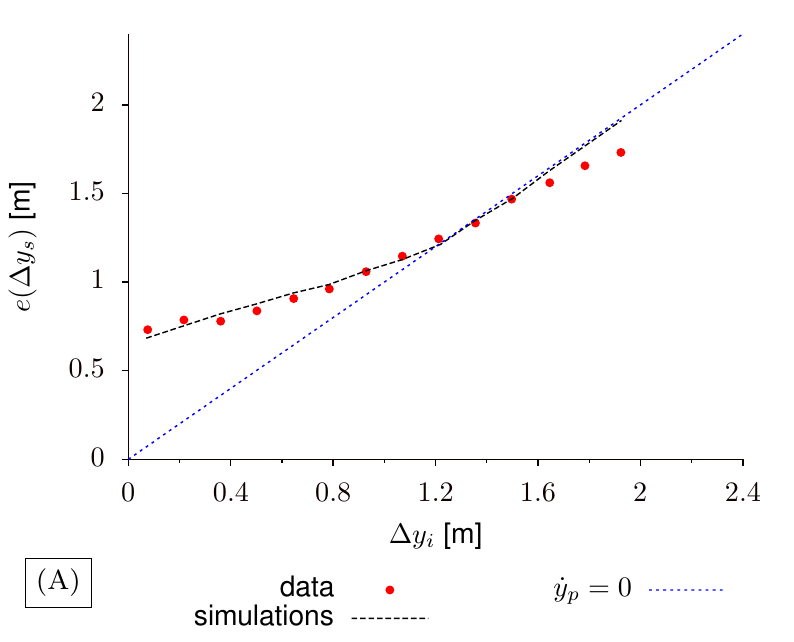}
\includegraphics[width=0.45\textwidth]{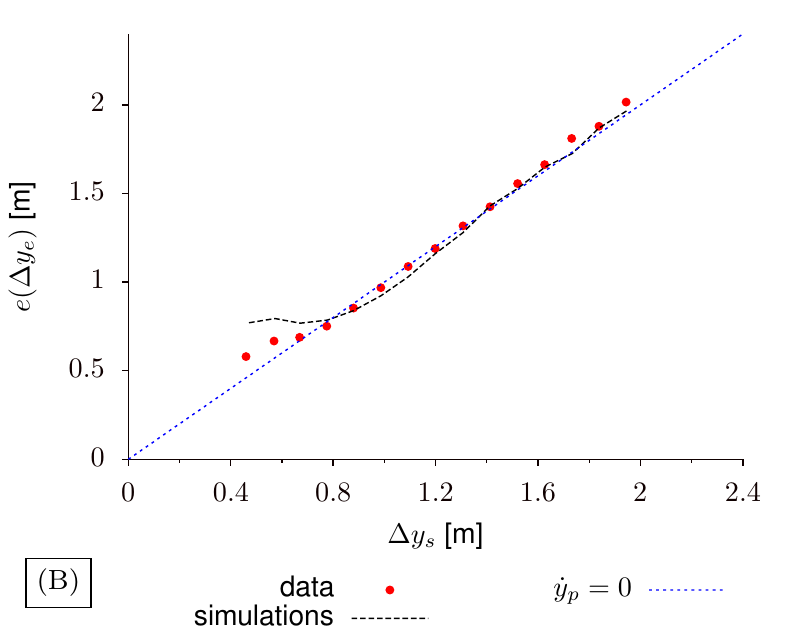}
  \caption{Average absolute lateral distance: comparison between data
    and simulations. (A) Absolute lateral distance when at the
    entrance ($|\Delta y_i|$, x-axis) vs. side-by-side (conditioned
    average $e(|\Delta y_s|)$, y-axis). (B) Absolute lateral
    distance when side-by-side ($|\Delta y_s|$, x-axis) vs. at the
    exit (conditioned average $e(|\Delta y_e|)$, y-axis).}         \label{fig:comparison-impactP}
\end{figure*}

For S1 (and analogously for S2), a
synthetic view of the data can be obtained by computing the ensemble
average of $|\Delta y_s|$ conditioned to $|\Delta y_i|$, namely
\begin{equation}\label{eq:ensavg-trend}
e(|\Delta y_s|) = \Eavg_{ens}\left[|\Delta y_s|\, \big|\, |\Delta y_i|\right] 
\end{equation}
Considering the approximation in Eq.~\eqref{eq:ensemble-approx-yp}, we expect this function to represent the deviation of the intended paths.
The scatter plots in Fig.~\ref{fig:comparison-IC}-\ref{fig:comparison-impactP} include data from $9089$ avoidance events (i.e. pairs of pedestrians) either experimentally measured (left panels) or simulated (right panels). 
In Fig.~\ref{fig:comparison-impactP}(A-B) we compare 
data and simulations in terms of the average
conditioned distances (Eq.~\eqref{eq:ensavg-trend}), respectively for
 scenarios S1 and S2.

\begin{figure*}[t]
\centering
\includegraphics[width=0.32\textwidth]{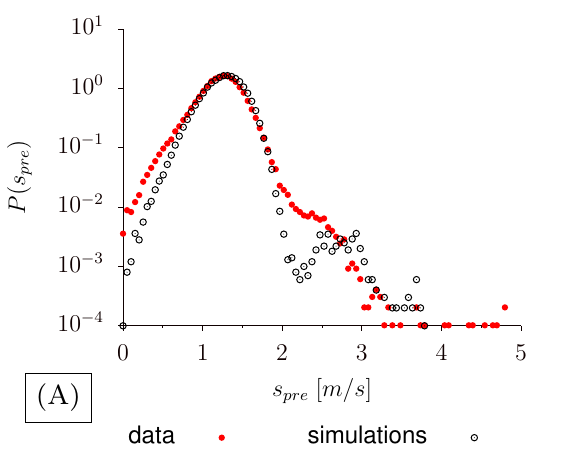}
\includegraphics[width=0.32\textwidth]{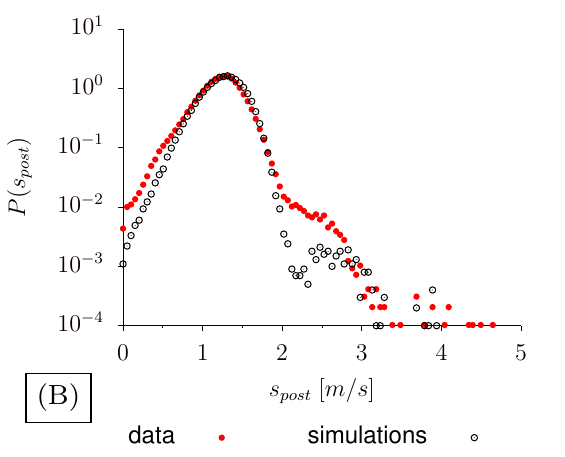}
\includegraphics[width=0.32\textwidth]{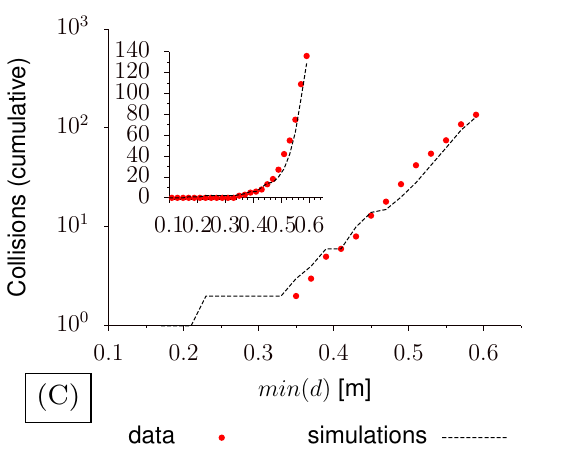}
  \caption{ (A, B) Speed distribution (i.e. distribution of the
    velocity modulus $s$) before and after the encounter.  Each time
    frame contributes one measurement per pedestrian. In (A) frames in
    the time interval $[t_s - 0.66\,\mbox{s}, t_s]$, where $t_s$ is
    the time instant of the side-by-side encounter, are considered. In (B) the
    considered time interval is $[t_s, t_s + 0.66\,\mbox{s}]$.  Note
    that for the pedestrians that are interacting, we have a reduced
    number of runners ($0.2\%$), and this is shown in the smaller
    percentage of speed values that are larger than $2\,$m/s. (C)
    Cumulative distribution of the minimum distance $\min(d)$ between
    the two pedestrians of the pair. Data and simulation are compared;
    the inset reports the same distribution in linear scale. This
    cumulative distribution expresses the number of collisions
    occurred vs. predicted. Effectively a collision happens for
    $\min(d)$ smaller or equal than the body diameter (about
    $0.5\,m$). Experimental data and simulations agree quantitatively on the
    exponential decaying trend.}  
       \label{fig:comparison-Collisions}
\end{figure*}

We observe the following
\begin{itemize}
\item[S1.] We expect avoidance maneuvers, especially when
  collision is imminent, i.e. for $|\Delta y_i| \lesssim s_b$, $s_b$
  being the size scale of the human body. In this condition, 
  we expect a modification of intended paths to yield
    $\ensavg{|\Delta y_s|} > \ensavg{|\Delta y_i|}$, which is consistent
  with our measurements in Fig.~\ref{fig:comparison-IC}(A). From the experimental measurements we have
  \begin{equation}
    \ensavg{|\Delta y_s|} \approx 0.75\,m\qquad\mbox{for $|\Delta y_i|\rightarrow 0$}
  \end{equation}
  and similarly in the case of simulations
  (Fig.~\ref{fig:comparison-IC}(B)). On the contrary, a decreasing
influence of $|\Delta y_i|$ on $|\Delta y_s|$ is expected as the
former increases, since no interaction is at play at large transversal
distances. As a consequence, we expect the average trend
\begin{equation}\label{eq:asympt-tr-deltay}
\ensavg{|\Delta y_s|} \approx \ensavg{|\Delta y_i|}   \ \mbox{for } |\Delta y_i| \gg s_b,
\end{equation}
i.e. a relaxation of $\ensavg{|\Delta y_s|}$ towards the diagonal of
the plane $(|\Delta y_i|, |\Delta y_s|)$. We observe such expected
trend (that we obtain in simulations per the scaling and anisotropy in
Eq.~\eqref{eq:Fvisiony}) only for $|\Delta y_i| < 1.4\,m$. In case
$|\Delta y_i| > 1.4\,m$, $\ensavg{|\Delta y_s|}$ lays slightly below
the diagonal line, suggesting an average ``end-distance contraction''.
As the distance increases, false positive and false
negative cases emerge in the selection operated by $\tilde{G}_{2,a}$. 
In this case, these determine the overweight
of the region below the diagonal in Fig.~\ref{fig:comparison-IC}(A),
thus the inflection of the $\ensavg{|\Delta y_s|}$ curve.

\item[S2.] The interaction dynamics in
  Fig.~\ref{fig:sketch-Trajectories} conjectures that the change in
  intended path mostly occurs to ensure avoidance, hence before the
  two pedestrians are closest in space. Afterwards it plays negligible
  role. In formulas the conjecture reads
\begin{equation}\label{eq:post-interaction-pref-path}
y_{p,1} - y_{p,2} \approx \mbox{const}\qquad\mbox{post-interaction}.
\end{equation}
Considering the approximation in Eq.~\eqref{eq:ensemble-approx-yp},
Eq.~\eqref{eq:post-interaction-pref-path} is consistent with our
measurements and simulations, reported in
Fig.~\ref{fig:comparison-CF}(A) and in
Fig.~\ref{fig:comparison-CF}(B), respectively. In these, it holds
\begin{equation}
\ensavg{|\Delta y_e|} \approx \ensavg{|\Delta y_s|}   \ \mbox{post-interaction},
\end{equation}
for $|\Delta y_s| > 0.8\,$m, while at lower $|\Delta y_s|$, i.e. for
very close passing distances (observed very rarely), we measure a small
tendency to increase the transversal distance in the post-interaction
stage. This can be an inertial phenomenon:  pedestrians
avoiding each other, yet passing by each other at very close distance,
keep on increasing their mutual distance. This aspect is modeled via the frontal
isotropic short range contact avoidance force in
Eq.~\eqref{eq:Fshortx}.
\end{itemize}
Avoidance impacts further on the walking speed $s=\sqrt{u^2+v^2}$, which we adopt as a
 second comparison term between model and experimental measurements.
Around the time instance $t_s$ of minimum distance (Fig.~\ref{fig:sketch-config}(ii)) 
speed is  temporarily
adjusted and reduced from the undisturbed flow regime
(Fig.~\ref{fig:PDF_Velocity_comparison}). Considering a time window
spanning $0.66\,$s before $t_s$ (i.e. $[t_s -0.66\,s, t_s]$) and a
time window spanning $0.66\,$s after $t_s$ (i.e. $[t_s, t_s +
  0.66\,s]$), in Fig.~\ref{fig:comparison-Collisions}(A) and
Fig.~\ref{fig:comparison-Collisions}(B), respectively, we compare the
speed distributions measured and predicted by the model. Also in this case we find excellent agreement.

Pairwise avoidance, finally, is not always a successful operation. As
a rare event, two persons can ``bump'' into each other, having their minimum
distance $\min(d)$ becoming comparable to their diameter. It is
important that such rare events remain captured -- in statistic terms -- by
the model, e.g. for their implication in safety. 
In Fig.~\ref{fig:comparison-Collisions}(C) we report the
cumulative distribution of ``collisions'' as a function of $\min(d)$. In
the range $[0.3,0.6]\,$m the distribution measured and the distribution  predicted by the 
model present an exponential growth in perfect agreement, with
about $40$ cases recorded in both cases for $\min(d) \leq 0.5\,$m.

\section{Discussion}
In this paper we investigated quantitatively 
the pedestrian-pedestrian avoidance
interactions occurring in diluted conditions comparing with measurements obtained in an
unprecedented real-life pedestrian tracking campaign.  As two
individuals walk in opposite directions on a shared space, avoidance
maneuvers become necessary should a collision be avoided - 
these maneuvers  affect, at the same time, the path observed and the intended path. We modeled
this scenario in terms of a sight-dependent interaction force and a collision-detering force, which
we superimposed to a Langevin model for the undisturbed pedestrian
motion. Overall the state of each individual was treated as a triple of variables 
including the components of the position plus a spatial variable representing the intended path.

We performed the experimental campaign employing state-of-the-art pedestraian tracking method, which in a period of over $6$ months, enabled us to collecte a dataset of about five million
high-resolution individual trajectories.
Using real-life data acquired from 24/7 tracking allows us, at the same time, to accurately quantify characteristic fluctuations and rare events
 in the dynamics (as events appearing only once in one thousand or once in ten thousand cases can be measured) and 
avoid potential biases related e.g. to the construction of laboratory/artificial experimental conditions (in which the dynamics to measure has been pre-defined and enforced by the experimenter).
Acquiring data in a real-life scenario, however, is somehow similar to
acquire data from different laboratory experiments, each having different experimental parameters,
that follow one another in random order. 
Ideally one wants to retain
only the measurements pertaining to the occurrences of a single virtual experiment (i.e. a single scenario) of interest, and aggregate them to perform ensemble statistical analyses. As operating a manual selection of measurements (as done in the past) would be 
impossible at the scale of our dataset, in the first half of the paper we propose a novel
method efficiently automatizing the selection. Representing the measurement through a graph we could
formally define scenarios of interest as well as efficiently identify them within the dataset. 
Selecting pairwise avoidance events in diluted conditions implies, in a naive prespective, a non-linear scan of the dataset: we are searching for pairs of trajectories that are mutually close while being far from any other at any time. 
With the method proposed a single pass of the dataset first, and a linear pass of the graph edges then, are sufficient
 to identify all the target events.

We analyzed the dynamics considering probability distribution functions that, thanks to our extensive dataset, 
result very well resolved even in the tails (rare events). 
Our models target and reproduce quantitatively the stochastic behavior observed. 
At the level of the undisturbed motion, considering a mixture of walkers and runners, we could reproduce the non-trivial
longitudinal dynamics
which shows fluctuations around two average speed values, one for walkers and one for runners, plus rare U-turning events.
This was possible via considering a double-well,
i.e. bi-stable, velocity potential for each of the two populations. 
The interplay by the white noise excitations 
and the gradient-type velocity dynamics
captures both small fluctuations around the average velocity, as well as
rare velocity inversions, that occur with a transition to the negative velocity stable state.

We addressed pairwise avoidance considering a social force-like interactions between undisturbed pedestrians, 
which we extended to affect an  hidden variable of the system introduced in this work: the individual intended path.
Despite hidden for single realizations, we believe that the variations in intended path can
 be measured on the basis of ensemble averages, that we computed on the transversal distances between the pedestrians.
Should pedestrians be in possible collision (initial transversal distance below $1.4\,$m), their intended paths are deflected, such that when passing by one another 
the mutual distance is no lower than $0.75\,$m. After the moment the passing occurs, no  further modifications in intended paths were recorded. By including the intended path in the dynamics and subjecting it to the social force, we could reproduce quantitatively the observed dynamics including speed reductions in the proximity of the passing as well as the number of collision events.

We consider this work as a first methodological and modeling step to treat quantitatively, in a statistically accurate sense, interactions in crowd dynamics. We believe that the present approach can be  extended to analyze situations characterized by higher complexity and density, increasingly common in civil infrastructures.

\begin{acknowledgments}
We acknowledge the Brilliant Streets research program of the Intelligent Lighting Institute at the
Eindhoven University of Technology, Nederlandse Spoorwegen, and the technical support of A. Muntean,
T. Kanters, A. Holten, G. Oerlemans and M. Speldenbrink. During the development of the infrastructure,
AC has been partly founded by a Lagrange Ph.D. scholarship granted by the CRT Foundation, Turin,
Italy and by the Eindhoven University of Technology, the Netherlands. This work is part of the JSTP
research programme “Vision driven visitor behaviour analysis and crowd management” with project
number 341-10-001, which is financed by the Netherlands Organisation for Scientific Research (NWO).
Support from COST Action MP1305 ``Flowing Matter'' is also
kindly acknowledged.
\end{acknowledgments}


\begin{figure*}[ht!]
\centering
  \includegraphics[width=\textwidth]{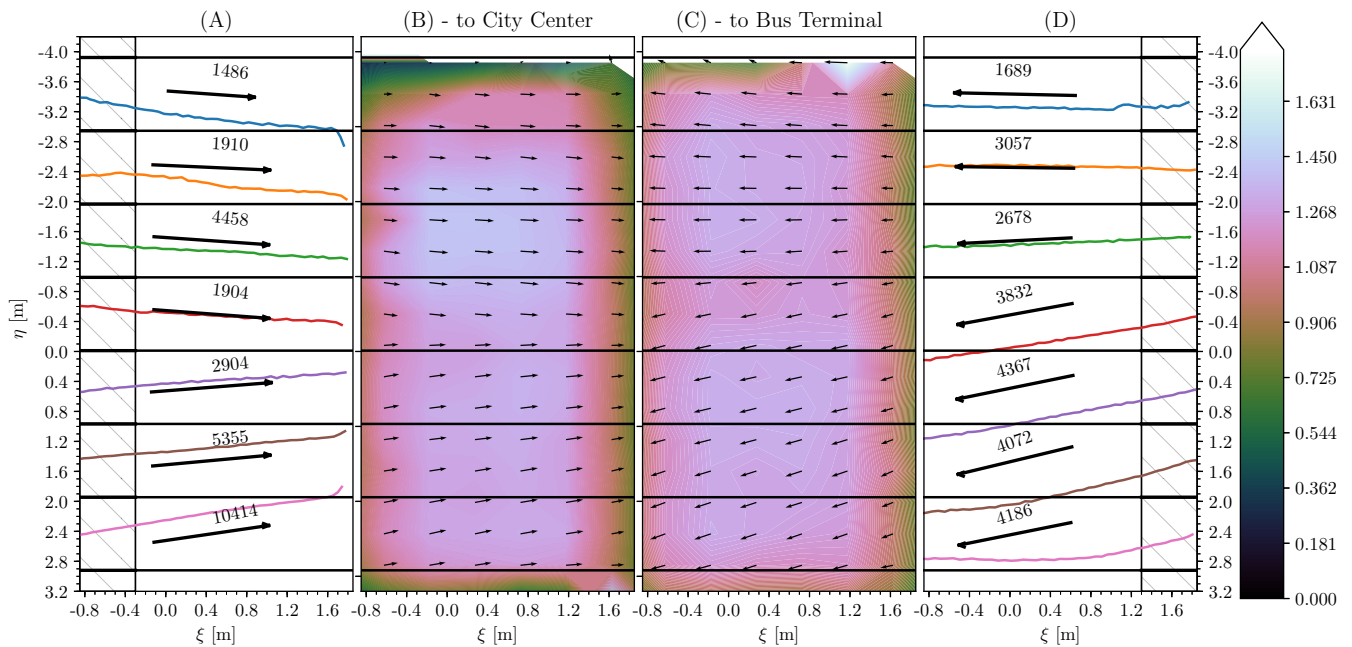}
  \caption{Average paths (A, D) and velocity (B, C) of
    undisturbed pedestrians going respectively to the city center
    (A, B) and to the the bus terminal (C, D) (cf. sketch in
    Fig.~\ref{fig:sketch-Eindhoven_Setup}(A)). Average paths (A, D) are
    computed by binning the trajectories by their initial point and
    then by averaging them in time (the number of trajectories per bin is reported). 
    Average velocities (B, C)  are computed employing a grid of 10x5 tiles. The backround
    colormap reports the average speed in m/s, according to the colorbar on the right.}  
       \label{fig:quiver_binned_statistic_2d}
\end{figure*}


\appendix
\section{Pedestrian tracking at Eindhoven Train Station: technicalities}\label{app:tracking}
Our data collection employed an array of four overhead \kinectTMS
sensors in order to obtain our depth-maps streams at VGA resolution
($640 \times 480$ pixels) and at $15\,$frames per second (fps). The
four views were in partial overlap and were merged into one large
canvas as those in Fig.~\ref{fig:sketch-graph-depth-maps}. The merging
algorithm and registration algorithms, treated
in~\cite{corbetta2016continuous}, rely on the fact that a depth map is
an overhead perspective view containing (by definition) the height of
each pixel. Such information is sufficient to obtain an axonometric
view (i.e. an aerial view in the limit of a far observer) from each of
the four streams. Considering a depth map via the cylindrical
coordinates $(\theta,\rho,h)$, where $\rho$ is a radial coordinate
(i.e. $\rho = 0$ is the image central axis), $\theta$ is an angle
spanning the image space around the image axis and $h$ is the altitude
from the ground (normalized such that $h=1$ is the floor and $h=0$ is
the camera plane), we employed the mapping
\begin{equation}\label{eq:coordinate-transf}
(\theta,\rho,h) \mapsto (\theta,\rho h,h)
\end{equation}
that displaces each point to its vertical line. Note that $\rho h \leq
\rho$ holds and $\rho h = \rho$ is true only at the ground level
(i.e. the ground level is invariant under this
transformation). Extracting the lowest depth value (top-most) for each
vertical line yields the desired axonometric view. Axonometric views
can be merged by simple superposition which requires just the
knowledge of the reciprocal positions of the cameras. Such
registration and calibration steps have been performed manually after
sliding a cart endowed with elements of known size underneath the
cameras (cf.~\cite{corbetta2016continuous} and appendix
in~\cite{corbetta2016multiscale}).

The four \kinectTMS sensors were connected in pairs to two computers whose system time has been synchronized with $O(1)\,$ms precision through \textit{Network Time Protocol} (NTP). To the best of our knowledge the moment at which a depth image is taken by a \kinectTMS is not controllable or triggerable. Hence, we let the four sensors record (maintaining $15\,$ fps), then we associated simultaneous images \textit{a posteriori}. This yields a maximum error of $33\,$ms i.e. approximately $3\,$cm considering an average walking speed of $1\,$m/s, i.e. less than $7$ pixels (where the conversion $230\,$px $\approx 1\,$m holds).

Merged depth images were processed to detect pedestrian positions via
the stochastic clustering algorithm proposed independently
in~\cite{seer2014kinects} and~\cite{brscic2013person}. We employed the
same implementation of our previous
works~\cite{corbetta2014TRP,corbetta2015MBE,corbetta2016fluctuations},
to which we refer for further details.

Differently from~\cite{corbetta2014TRP,corbetta2015MBE,corbetta2016fluctuations}, we revised the tracking approach
for increased robustness. As
in~\cite{corbetta2014TRP,corbetta2015MBE,corbetta2016fluctuations}
pedestrians are tracked employing the classical particle tracking
velocimetry (PTV) approach (e.g.~\cite{willneff2003spatio}) and
specifically through the OpenPTV library~\cite{OpenPTV}. However each
detected pedestrian is considered via five different points:
\begin{inparaenum}
\item the centroid of the body;
\item the estimated head top  position (centroid of the points within the $5^{th}$ depth percentile);
\item the estimated head-neck position (centroid of the points within the $10^{th}$ depth percentile);
\item the estimated head-neck-torso centroid (centroid of the points within the $20^{th}$ depth percentile);
\item the upper half of the body (centroid of the points within the $50^{th}$ depth percentile).
\end{inparaenum}

We performed five independent tracking considering only one of these
five positions at a time and for all pedestrians. Hence, we compared
the five tracking results for consistency considering reliable the
tracks for which at least $3$ out of $5$ tracking results were in
agreement.

We smoothed the obtained trajectories for noise reduction with the
Savisky-Golay algorithm~\cite{savitzky1964smoothing}, a common
approach in PTV (e.g.~\cite{PTV-liberzon}), with window size $5$ and
polynomial degree $2$.

\section{Coordinate transformation for undisturbed pedestrian trajectories}\label{app:coord-single}
In Fig.~\ref{fig:quiver_binned_statistic_2d} we include different
average measurements of the undisturbed motion. We observe that
pedestrian trajectories although generally straight, are not parallel
to the axis of the station walkway (i.e. the horizontal direction in
Fig.~\ref{fig:quiver_binned_statistic_2d}). To employ the $x,y$
coordinate system introduced in Sect.~\ref{sect:diluted}, we rotate
each trajectory to have the longitudinal direction aligned with the
$x$ axis. To this aim, we perform a $2^{nd}$ order polynomial fitting
of the two components of each trajectory (in the horizontal-vertical
reference of Fig.~\ref{fig:quiver_binned_statistic_2d}, say
$(\xi,\eta)$) as a function of time. In formulas, we fit
\begin{equation*}
t \mapsto (\xi,\eta)
\end{equation*}
via
\begin{equation}\label{eq:trajfit}
t \mapsto (a_\xi t^2 + b_\xi t + c_\xi, a_\eta t^2 + b_\eta t + c_\eta).
\end{equation}
Hence, we rotate each trajectory to align the tangent line of the polynomial fit at $t=0$ to the $x$ axis, i.e. 
\begin{equation*}
(x,y)^T = \frac{1}{b_\eta^2 + b_\xi^2}
\left(\begin{matrix}
b_\xi & b_\eta \\
- b_\eta &  b_\xi
\end{matrix}\right)(\xi,\eta)^T. 
\end{equation*}
Performing such a transformation, trajectories that are straight are
simply aligned to the $x$ axis. In turn, U-shaped trajectories, the
most common example of inversion dynamics, get just their first
(and possibly last) portions aligned with the $x$ axis.

\begin{figure}[ht!]
\centering
\includegraphics[width=.20\textwidth]{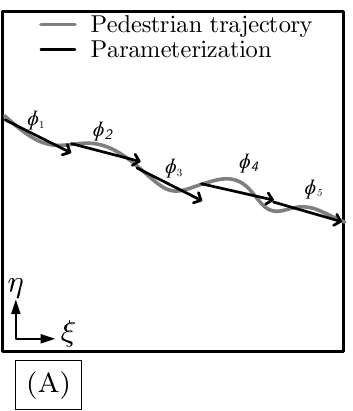}
\includegraphics[width=.27\textwidth]{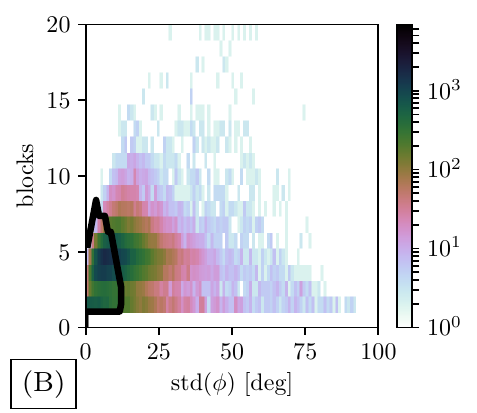}
  \caption{(A) The analysis of the diluted motion assumes the
    trajectory to be a fluctuation around a straight intended path,
    which we call ``quasi-rectilinear''. In order to isolate
    quasi-rectilinear trajectories, each track is divided into sets of
    $7$ contiguous frames (e.g. indexed by $i$). For each set the
    angle $\phi_i$ between the average velocity and the longitudinal
    direction in the corridor ($\xi$) is evaluated. Quasi-rectilinear
    trajectories (as in the panel) feature low variance within the set
    $\{\phi_i\}$.  Trajectories exhibiting instead significant drifts
    (high curvature, i.e. high $\{\phi_i\}$ variance) are neglected.
      (B) Joint distribution of $\mbox{std}(\phi)$ ($x$-axis,
    cf. Eq.~\eqref{eq:angle-phi-per-block}) and the number of blocks
    ($y$-axis, one block gathers $7$ consecutive frames) for
    undisturbed pedestrians. The number of measurements per bin ($1^o
    \times 1\,$block) follows the logarithmic colorbar. The region
    marked identifies the trajectories employed.}  
 \label{fig:hist2d_angle}
\end{figure}

\paragraph{Quasi-rectilinear trajectories.} 
We use a similar fitting approach to distinguish trajectories that are
quasi-rectilinear (straight but small fluctuations) from trajectories
that exhibit large curvature or even turn-backs. We split each
trajectory in blocks of $7\,$ frames (i.e. about half a second,
starting from the first frame), that we independently fit as in
Eq.~\eqref{eq:trajfit}. See Fig.~\ref{fig:hist2d_angle}(A).  In case the last block of a trajectory
contains less than $7\,$frames we neglect it. We estimate the angular
slope of each block with reference to the horizontal direction as
\begin{equation}\label{eq:angle-phi-per-block}
\phi = |\arctan(b_\eta/b_\xi)|.
\end{equation}
Rectilinear trajectories satisfy $\mbox{std}(\phi) \approx 0^+$, where
we evaluated the standard deviation ($\mbox{std}$) on a block
basis. In Fig.~\ref{fig:hist2d_angle}(B) we report the joint distribution
of $\mbox{std}(\phi)$ and the number of blocks per trajectory (i.e. a
measure of the trajectory length). Neglecting the very short
trajectories ($1\,$ block case), we notice that most of our
measurements lay in the low-$\mbox{std}(\phi)$ region having between
$2$ and $6$ blocks. These trajectories are mostly straight and
encompass normal walking velocities. These feature Gaussian
transversal fluctuations as discussed in
Fig.~\ref{fig:PDF_Velocity_comparison}(B).

\section{Coordinate transformation for pedestrian pairs in avoidance}\label{app:coord-pairs}
\begin{figure}[ht!]
 \centering
  \includegraphics[width=.35\textwidth]{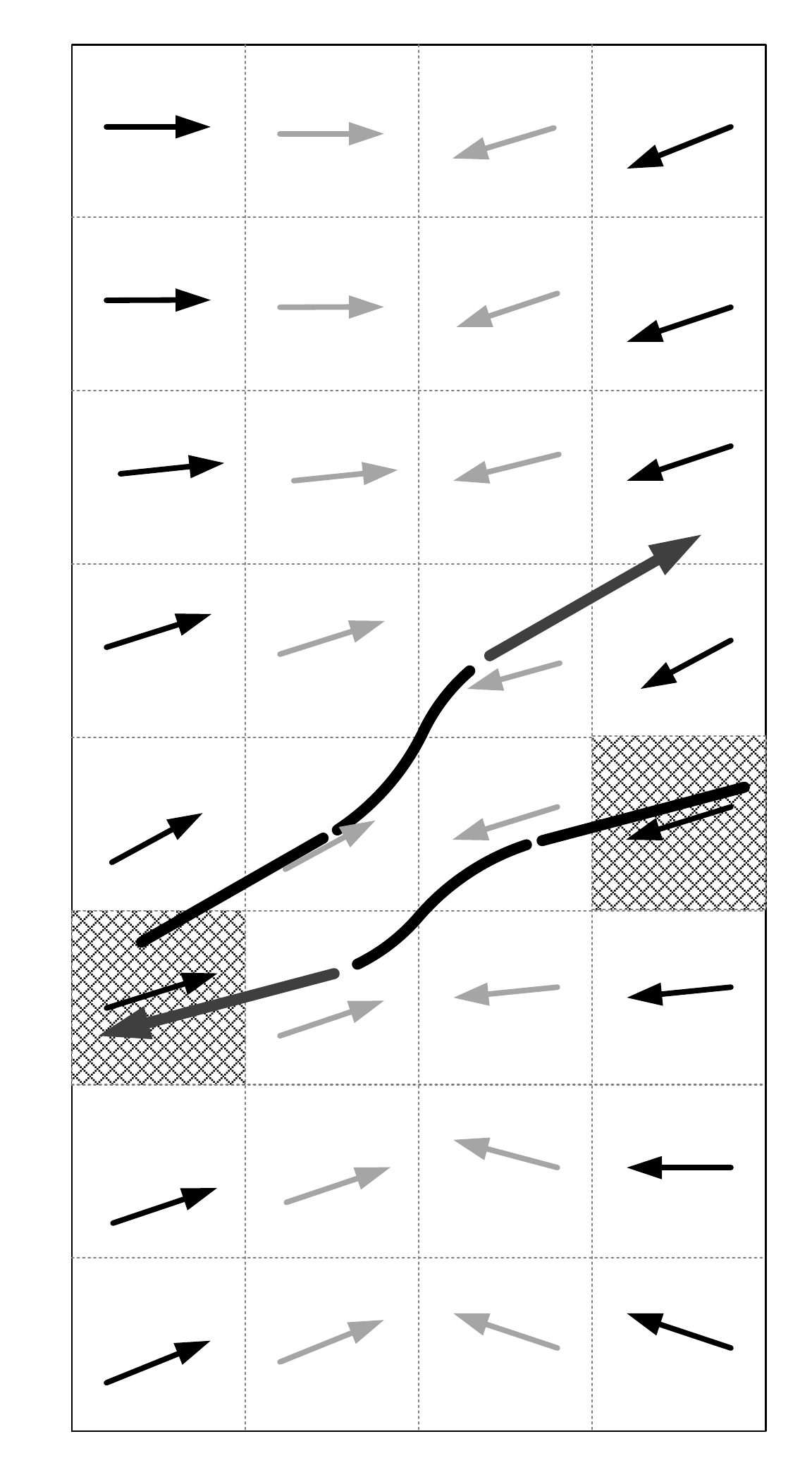}
  \caption{Coordinate transformation for pedestrian pairs in avoidance
    for analysis convenience. Trajectories get rotated around the
    entrance point to compensate for intended paths not parallel to the
    corridor axis and then individually translated to conserve the
    minimum pedestrian distance. Cf. the algorithm in
    Appendix~\ref{app:coord-pairs}.} 
           \label{fig:Pair_trajectory_correction}
\end{figure}

In the analysis of pairwise avoidance interaction we performed a
coordinate transformation, in similar spirit to
Appendix~\ref{app:coord-single}, to bring the pairs of trajectories to a
coordinate system convenient for the analysis and removed of average
motions. The rationale for the transformation followed (i) the minimum distance between pedestrian cannot be
altered; (ii) the intended path of a pedestrian entering in a given
position is the same of an undisturbed pedestrian entering in the same
position. The deviations from such intended paths are what determines
the $\Delta y$ variables. In algorithmic terms we proceeded as
follows:
\begin{enumerate}
\item given a $(\xi,\eta)$  grid (coarse)  as in
  Fig.~\ref{fig:Pair_trajectory_correction}, we calculated the average
  motion of undisturbed pedestrians directed both to the city center
  and to the bus terminal. Let $\theta^{1,C}(\xi,\eta)$ and
  $\theta^{1,T}(\xi,\eta)$ be the angle of the average velocity with the
  longitudinal axis of the corridor ($\xi$ direction).
\item trajectories for pairs in interactions are rotated around their
  entering point $(\xi_0,\eta_0)$ of an angle $-\theta^{1,C}(\xi_0,\eta_0)$ or
  $-\theta^{1,T}(\xi_0,\eta_0)$ (in dependence on the direction). This
  compensates for intended paths that are straight but not parallel to
  the $\xi$ axis to respect (ii).
\item trajectories are translated apart to respect (i).
\end{enumerate}
The quantities $\Delta y_i$, $\Delta y_s$, $\Delta y_e$ are finally
calculated after these roto-translations.

\section{Sensitivity of the graph-based selection}
\begin{figure}[ht!]
 \centering
\includegraphics[width=.23\textwidth]{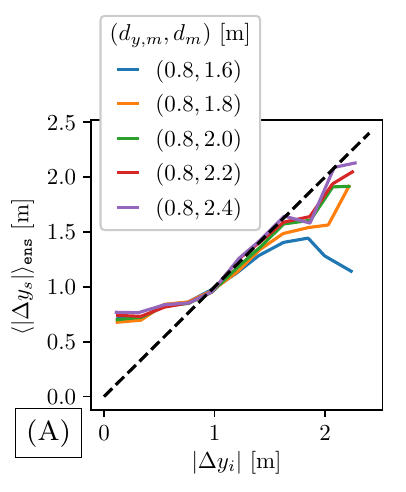}
\includegraphics[width=.23\textwidth]{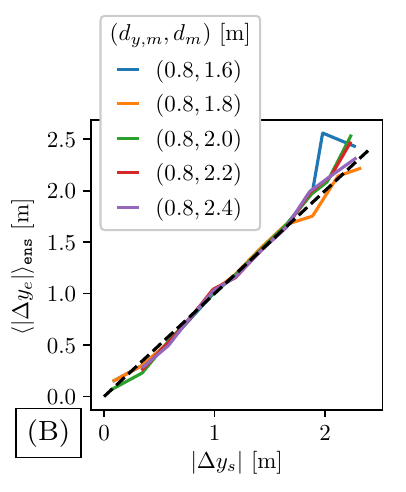}
  \caption{Average transversal distance for cases S1 (A) and S2 (B) 
    (cf. Sect.~\ref{sec:interactions}) as a function of the parameter $d_m$
    used to sparsify the graph $G$. In case S1 (A) a dependence on $d_m$
    can be observed for $|\Delta y_i| > 1\,$m. Notably and consistently with the points (ii)-(iii) in
    Sect.~\ref{sec:interactions}, as the region $\intGrI_p$ grows in
    size, i.e. more and more pairs are included, the asymptotic
    behavior in Eq.~\eqref{eq:asympt-tr-deltay} is recovered with
    increased accuracy. In case S2 (B) no particular dependency on the
    parameters is observed.}        \label{fig:selection_parameter_study_IC}
\end{figure}

The parameters that we employed to sparsify the graph $G$, and select pairs
of pedestrians avoiding each other, have been chosen considering a
typical size scale of the interaction. Yet, it is reasonable to expect
a sensitivity of the distributions in Fig.~\ref{fig:comparison-IC}(A) and
Fig.~\ref{fig:comparison-CF}(A) with respect to these parameters.
This sensitivity appears minimum, considering, for instance, average trends from
Eq.~\eqref{eq:ensavg-trend}. In
Fig.~\ref{fig:selection_parameter_study_IC}(A)
and~\ref{fig:selection_parameter_study_IC}(B), corresponding
respectively to Fig.~\ref{fig:comparison-IC}(A) and
Fig.~\ref{fig:comparison-CF}(A), we plot Eq.~\eqref{eq:ensavg-trend}
in dependence of the parameters $d_m$ and $d_{y,m}$. For computational
reasons the data are restricted to only one day of measurements
($27^{th}$ November 2014).

\newpage
\bibliography{./master}

\begin{thebibliography}{37}%
\makeatletter
\providecommand \@ifxundefined [1]{%
 \@ifx{#1\undefined}
}%
\providecommand \@ifnum [1]{%
 \ifnum #1\expandafter \@firstoftwo
 \else \expandafter \@secondoftwo
 \fi
}%
\providecommand \@ifx [1]{%
 \ifx #1\expandafter \@firstoftwo
 \else \expandafter \@secondoftwo
 \fi
}%
\providecommand \natexlab [1]{#1}%
\providecommand \enquote  [1]{``#1''}%
\providecommand \bibnamefont  [1]{#1}%
\providecommand \bibfnamefont [1]{#1}%
\providecommand \citenamefont [1]{#1}%
\providecommand \href@noop [0]{\@secondoftwo}%
\providecommand \href [0]{\begingroup \@sanitize@url \@href}%
\providecommand \@href[1]{\@@startlink{#1}\@@href}%
\providecommand \@@href[1]{\endgroup#1\@@endlink}%
\providecommand \@sanitize@url [0]{\catcode `\\12\catcode `\$12\catcode
  `\&12\catcode `\#12\catcode `\^12\catcode `\_12\catcode `\%12\relax}%
\providecommand \@@startlink[1]{}%
\providecommand \@@endlink[0]{}%
\providecommand \url  [0]{\begingroup\@sanitize@url \@url }%
\providecommand \@url [1]{\endgroup\@href {#1}{\urlprefix }}%
\providecommand \urlprefix  [0]{URL }%
\providecommand \Eprint [0]{\href }%
\providecommand \doibase [0]{http://dx.doi.org/}%
\providecommand \selectlanguage [0]{\@gobble}%
\providecommand \bibinfo  [0]{\@secondoftwo}%
\providecommand \bibfield  [0]{\@secondoftwo}%
\providecommand \translation [1]{[#1]}%
\providecommand \BibitemOpen [0]{}%
\providecommand \bibitemStop [0]{}%
\providecommand \bibitemNoStop [0]{.\EOS\space}%
\providecommand \EOS [0]{\spacefactor3000\relax}%
\providecommand \BibitemShut  [1]{\csname bibitem#1\endcsname}%
\let\auto@bib@innerbib\@empty
\bibitem [{\citenamefont {Cristiani}\ \emph {et~al.}(2014)\citenamefont
  {Cristiani}, \citenamefont {Piccoli},\ and\ \citenamefont
  {Tosin}}]{cristiani2014BOOK}%
  \BibitemOpen
  \bibfield  {author} {\bibinfo {author} {\bibfnamefont {E.}~\bibnamefont
  {Cristiani}}, \bibinfo {author} {\bibfnamefont {B.}~\bibnamefont {Piccoli}},
  \ and\ \bibinfo {author} {\bibfnamefont {A.}~\bibnamefont {Tosin}},\
  }\href@noop {} {\emph {\bibinfo {title} {Multiscale {M}odeling of
  {P}edestrian {D}ynamics}}},\ \bibinfo {series} {Modeling, Simulation and
  Applications}, Vol.~\bibinfo {volume} {12}\ (\bibinfo  {publisher}
  {Springer},\ \bibinfo {year} {2014})\BibitemShut {NoStop}%
\bibitem [{\citenamefont {Moussa{\"\i}d}\ \emph {et~al.}(2011)\citenamefont
  {Moussa{\"\i}d}, \citenamefont {Helbing},\ and\ \citenamefont
  {Theraulaz}}]{Mou-pnas}%
  \BibitemOpen
  \bibfield  {author} {\bibinfo {author} {\bibfnamefont {M.}~\bibnamefont
  {Moussa{\"\i}d}}, \bibinfo {author} {\bibfnamefont {D.}~\bibnamefont
  {Helbing}}, \ and\ \bibinfo {author} {\bibfnamefont {G.}~\bibnamefont
  {Theraulaz}},\ }\href@noop {} {\bibfield  {journal} {\bibinfo  {journal} {P.
  Natl. Acad. Sci. Usa}\ }\textbf {\bibinfo {volume} {108}},\ \bibinfo {pages}
  {6884} (\bibinfo {year} {2011})}\BibitemShut {NoStop}%
\bibitem [{\citenamefont {Moussa{\"\i}d}\ \emph
  {et~al.}(2009{\natexlab{a}})\citenamefont {Moussa{\"\i}d}, \citenamefont
  {Garnier}, \citenamefont {Theraulaz},\ and\ \citenamefont
  {Helbing}}]{moussaid2009collective}%
  \BibitemOpen
  \bibfield  {author} {\bibinfo {author} {\bibfnamefont {M.}~\bibnamefont
  {Moussa{\"\i}d}}, \bibinfo {author} {\bibfnamefont {S.}~\bibnamefont
  {Garnier}}, \bibinfo {author} {\bibfnamefont {G.}~\bibnamefont {Theraulaz}},
  \ and\ \bibinfo {author} {\bibfnamefont {D.}~\bibnamefont {Helbing}},\
  }\href@noop {} {\bibfield  {journal} {\bibinfo  {journal} {Top. Cogn. Sci.}\
  }\textbf {\bibinfo {volume} {1}},\ \bibinfo {pages} {469} (\bibinfo {year}
  {2009}{\natexlab{a}})}\BibitemShut {NoStop}%
\bibitem [{\citenamefont {Moussa{\"\i}d}\ \emph
  {et~al.}(2009{\natexlab{b}})\citenamefont {Moussa{\"\i}d}, \citenamefont
  {Helbing}, \citenamefont {Garnier}, \citenamefont {Johansson}, \citenamefont
  {Combe},\ and\ \citenamefont {Theraulaz}}]{Moussadrspb.2009.0405}%
  \BibitemOpen
  \bibfield  {author} {\bibinfo {author} {\bibfnamefont {M.}~\bibnamefont
  {Moussa{\"\i}d}}, \bibinfo {author} {\bibfnamefont {D.}~\bibnamefont
  {Helbing}}, \bibinfo {author} {\bibfnamefont {S.}~\bibnamefont {Garnier}},
  \bibinfo {author} {\bibfnamefont {A.}~\bibnamefont {Johansson}}, \bibinfo
  {author} {\bibfnamefont {M.}~\bibnamefont {Combe}}, \ and\ \bibinfo {author}
  {\bibfnamefont {G.}~\bibnamefont {Theraulaz}},\ }\href {\doibase
  10.1098/rspb.2009.0405} {\bibfield  {journal} {\bibinfo  {journal} {Proc. R.
  Soc. Lond., B, Biol. Sci.}\ } (\bibinfo {year} {2009}{\natexlab{b}}),\
  10.1098/rspb.2009.0405}\BibitemShut {NoStop}%
\bibitem [{\citenamefont {Romanczuk}\ \emph {et~al.}(2012)\citenamefont
  {Romanczuk}, \citenamefont {B{\"a}r}, \citenamefont {Ebeling}, \citenamefont
  {Lindner},\ and\ \citenamefont {Schimansky-Geier}}]{Lutz}%
  \BibitemOpen
  \bibfield  {author} {\bibinfo {author} {\bibfnamefont {P.}~\bibnamefont
  {Romanczuk}}, \bibinfo {author} {\bibfnamefont {M.}~\bibnamefont {B{\"a}r}},
  \bibinfo {author} {\bibfnamefont {W.}~\bibnamefont {Ebeling}}, \bibinfo
  {author} {\bibfnamefont {B.}~\bibnamefont {Lindner}}, \ and\ \bibinfo
  {author} {\bibfnamefont {L.}~\bibnamefont {Schimansky-Geier}},\ }\href@noop
  {} {\bibfield  {journal} {\bibinfo  {journal} {Eur. Phys. J. Special Topics}\
  }\textbf {\bibinfo {volume} {202}},\ \bibinfo {pages} {1} (\bibinfo {year}
  {2012})}\BibitemShut {NoStop}%
\bibitem [{\citenamefont {Bellomo}\ \emph {et~al.}(2012)\citenamefont
  {Bellomo}, \citenamefont {Piccoli},\ and\ \citenamefont
  {Tosin}}]{bellomo2012modeling}%
  \BibitemOpen
  \bibfield  {author} {\bibinfo {author} {\bibfnamefont {N.}~\bibnamefont
  {Bellomo}}, \bibinfo {author} {\bibfnamefont {B.}~\bibnamefont {Piccoli}}, \
  and\ \bibinfo {author} {\bibfnamefont {A.}~\bibnamefont {Tosin}},\
  }\href@noop {} {\bibfield  {journal} {\bibinfo  {journal} {Mathematical
  Models and Methods in Applied Sciences}\ }\textbf {\bibinfo {volume} {22}},\
  \bibinfo {pages} {1230004} (\bibinfo {year} {2012})}\BibitemShut {NoStop}%
\bibitem [{\citenamefont {Wagoum}\ \emph {et~al.}(2012)\citenamefont {Wagoum},
  \citenamefont {Seyfried},\ and\ \citenamefont {Holl}}]{Sey1}%
  \BibitemOpen
  \bibfield  {author} {\bibinfo {author} {\bibfnamefont {A.~U.~K.}\
  \bibnamefont {Wagoum}}, \bibinfo {author} {\bibfnamefont {A.}~\bibnamefont
  {Seyfried}}, \ and\ \bibinfo {author} {\bibfnamefont {S.}~\bibnamefont
  {Holl}},\ }\href@noop {} {\bibfield  {journal} {\bibinfo  {journal} {Advances
  in Complex Systems}\ }\textbf {\bibinfo {volume} {15}},\ \bibinfo {pages}
  {1250029} (\bibinfo {year} {2012})}\BibitemShut {NoStop}%
\bibitem [{\citenamefont {Duives}\ \emph {et~al.}(2013)\citenamefont {Duives},
  \citenamefont {Daamen},\ and\ \citenamefont {Hoogendoorn}}]{Duives}%
  \BibitemOpen
  \bibfield  {author} {\bibinfo {author} {\bibfnamefont {D.~C.}\ \bibnamefont
  {Duives}}, \bibinfo {author} {\bibfnamefont {W.}~\bibnamefont {Daamen}}, \
  and\ \bibinfo {author} {\bibfnamefont {S.~P.}\ \bibnamefont {Hoogendoorn}},\
  }\href@noop {} {\bibfield  {journal} {\bibinfo  {journal} {Transportation
  Research Part C: Emerging Technologies}\ }\textbf {\bibinfo {volume} {37}},\
  \bibinfo {pages} {193 } (\bibinfo {year} {2013})}\BibitemShut {NoStop}%
\bibitem [{\citenamefont {Hughes}(2003)}]{hughes2003flow}%
  \BibitemOpen
  \bibfield  {author} {\bibinfo {author} {\bibfnamefont {R.~L.}\ \bibnamefont
  {Hughes}},\ }\href@noop {} {\bibfield  {journal} {\bibinfo  {journal} {Annual
  Review of Fluid Mechanics}\ }\textbf {\bibinfo {volume} {35}},\ \bibinfo
  {pages} {169} (\bibinfo {year} {2003})}\BibitemShut {NoStop}%
\bibitem [{\citenamefont {Helbing}\ and\ \citenamefont
  {Moln\'ar}(1995)}]{helbing1995PRE}%
  \BibitemOpen
  \bibfield  {author} {\bibinfo {author} {\bibfnamefont {D.}~\bibnamefont
  {Helbing}}\ and\ \bibinfo {author} {\bibfnamefont {P.}~\bibnamefont
  {Moln\'ar}},\ }\href {\doibase 10.1103/PhysRevE.51.4282} {\bibfield
  {journal} {\bibinfo  {journal} {Phys. Rev. E}\ }\textbf {\bibinfo {volume}
  {51}},\ \bibinfo {pages} {4282} (\bibinfo {year} {1995})}\BibitemShut
  {NoStop}%
\bibitem [{\citenamefont {Marchetti}\ \emph {et~al.}(2013)\citenamefont
  {Marchetti}, \citenamefont {Joanny}, \citenamefont {Ramaswamy}, \citenamefont
  {Liverpool}, \citenamefont {Prost}, \citenamefont {Rao},\ and\ \citenamefont
  {Simha}}]{RevModPhys.85.1143}%
  \BibitemOpen
  \bibfield  {author} {\bibinfo {author} {\bibfnamefont {M.~C.}\ \bibnamefont
  {Marchetti}}, \bibinfo {author} {\bibfnamefont {J.~F.}\ \bibnamefont
  {Joanny}}, \bibinfo {author} {\bibfnamefont {S.}~\bibnamefont {Ramaswamy}},
  \bibinfo {author} {\bibfnamefont {T.~B.}\ \bibnamefont {Liverpool}}, \bibinfo
  {author} {\bibfnamefont {J.}~\bibnamefont {Prost}}, \bibinfo {author}
  {\bibfnamefont {M.}~\bibnamefont {Rao}}, \ and\ \bibinfo {author}
  {\bibfnamefont {R.~A.}\ \bibnamefont {Simha}},\ }\href {\doibase
  10.1103/RevModPhys.85.1143} {\bibfield  {journal} {\bibinfo  {journal} {Rev.
  Mod. Phys.}\ }\textbf {\bibinfo {volume} {85}},\ \bibinfo {pages} {1143}
  (\bibinfo {year} {2013})}\BibitemShut {NoStop}%
\bibitem [{\citenamefont {Corbetta}\ \emph
  {et~al.}(2017{\natexlab{a}})\citenamefont {Corbetta}, \citenamefont {Lee},
  \citenamefont {Benzi}, \citenamefont {Muntean},\ and\ \citenamefont
  {Toschi}}]{corbetta2016fluctuations}%
  \BibitemOpen
  \bibfield  {author} {\bibinfo {author} {\bibfnamefont {A.}~\bibnamefont
  {Corbetta}}, \bibinfo {author} {\bibfnamefont {C.}~\bibnamefont {Lee}},
  \bibinfo {author} {\bibfnamefont {R.}~\bibnamefont {Benzi}}, \bibinfo
  {author} {\bibfnamefont {A.}~\bibnamefont {Muntean}}, \ and\ \bibinfo
  {author} {\bibfnamefont {F.}~\bibnamefont {Toschi}},\ }\href@noop {}
  {\bibfield  {journal} {\bibinfo  {journal} {Phys. Rev. E}\ }\textbf {\bibinfo
  {volume} {95}},\ \bibinfo {pages} {032316} (\bibinfo {year}
  {2017}{\natexlab{a}})}\BibitemShut {NoStop}%
\bibitem [{\citenamefont {Moussa{\"\i}d}\ \emph {et~al.}(2010)\citenamefont
  {Moussa{\"\i}d}, \citenamefont {Perozo}, \citenamefont {Garnier},
  \citenamefont {Helbing},\ and\ \citenamefont
  {Theraulaz}}]{10.1371/journal.pone.0010047}%
  \BibitemOpen
  \bibfield  {author} {\bibinfo {author} {\bibfnamefont {M.}~\bibnamefont
  {Moussa{\"\i}d}}, \bibinfo {author} {\bibfnamefont {N.}~\bibnamefont
  {Perozo}}, \bibinfo {author} {\bibfnamefont {S.}~\bibnamefont {Garnier}},
  \bibinfo {author} {\bibfnamefont {D.}~\bibnamefont {Helbing}}, \ and\
  \bibinfo {author} {\bibfnamefont {G.}~\bibnamefont {Theraulaz}},\ }\href
  {\doibase 10.1371/journal.pone.0010047} {\bibfield  {journal} {\bibinfo
  {journal} {PLoS ONE}\ }\textbf {\bibinfo {volume} {5}},\ \bibinfo {pages} {1}
  (\bibinfo {year} {2010})}\BibitemShut {NoStop}%
\bibitem [{\citenamefont {Corbetta}\ \emph
  {et~al.}(2016{\natexlab{a}})\citenamefont {Corbetta}, \citenamefont
  {Meeusen}, \citenamefont {Lee},\ and\ \citenamefont
  {Toschi}}]{corbetta2016continuous}%
  \BibitemOpen
  \bibfield  {author} {\bibinfo {author} {\bibfnamefont {A.}~\bibnamefont
  {Corbetta}}, \bibinfo {author} {\bibfnamefont {J.}~\bibnamefont {Meeusen}},
  \bibinfo {author} {\bibfnamefont {C.}~\bibnamefont {Lee}}, \ and\ \bibinfo
  {author} {\bibfnamefont {F.}~\bibnamefont {Toschi}},\ }in\ \href@noop {}
  {\emph {\bibinfo {booktitle} {Pedestrian and Evacuation Dynamics 2016}}}\
  (\bibinfo  {publisher} {University of Science and Technology of China
  press},\ \bibinfo {year} {2016})\ pp.\ \bibinfo {pages} {18--24}\BibitemShut
  {NoStop}%
\bibitem [{\citenamefont {Kretz}\ \emph {et~al.}(2006)\citenamefont {Kretz},
  \citenamefont {Gr{\"u}nebohm}, \citenamefont {Kaufman}, \citenamefont
  {Mazur},\ and\ \citenamefont {Schreckenberg}}]{kretz2006experimental}%
  \BibitemOpen
  \bibfield  {author} {\bibinfo {author} {\bibfnamefont {T.}~\bibnamefont
  {Kretz}}, \bibinfo {author} {\bibfnamefont {A.}~\bibnamefont
  {Gr{\"u}nebohm}}, \bibinfo {author} {\bibfnamefont {M.}~\bibnamefont
  {Kaufman}}, \bibinfo {author} {\bibfnamefont {F.}~\bibnamefont {Mazur}}, \
  and\ \bibinfo {author} {\bibfnamefont {M.}~\bibnamefont {Schreckenberg}},\
  }\href@noop {} {\bibfield  {journal} {\bibinfo  {journal} {J. Stat.
  Mech.-Theory E.}\ }\textbf {\bibinfo {volume} {2006}},\ \bibinfo {pages}
  {P10001} (\bibinfo {year} {2006})}\BibitemShut {NoStop}%
\bibitem [{\citenamefont {Zhang}\ and\ \citenamefont
  {Seyfried}(2014)}]{6957746}%
  \BibitemOpen
  \bibfield  {author} {\bibinfo {author} {\bibfnamefont {J.}~\bibnamefont
  {Zhang}}\ and\ \bibinfo {author} {\bibfnamefont {A.}~\bibnamefont
  {Seyfried}},\ }in\ \href {\doibase 10.1109/ITSC.2014.6957746} {\emph
  {\bibinfo {booktitle} {17th International IEEE Conference on Intelligent
  Transportation Systems (ITSC)}}}\ (\bibinfo {year} {2014})\ pp.\ \bibinfo
  {pages} {542--547}\BibitemShut {NoStop}%
\bibitem [{\citenamefont {Zanlungo}\ \emph {et~al.}(2014)\citenamefont
  {Zanlungo}, \citenamefont {Ikeda},\ and\ \citenamefont
  {Kanda}}]{PhysRevE.89.012811}%
  \BibitemOpen
  \bibfield  {author} {\bibinfo {author} {\bibfnamefont {F.}~\bibnamefont
  {Zanlungo}}, \bibinfo {author} {\bibfnamefont {T.}~\bibnamefont {Ikeda}}, \
  and\ \bibinfo {author} {\bibfnamefont {T.}~\bibnamefont {Kanda}},\ }\href
  {\doibase 10.1103/PhysRevE.89.012811} {\bibfield  {journal} {\bibinfo
  {journal} {Phys. Rev. E}\ }\textbf {\bibinfo {volume} {89}},\ \bibinfo
  {pages} {012811} (\bibinfo {year} {2014})}\BibitemShut {NoStop}%
\bibitem [{\citenamefont {Tamura}\ \emph {et~al.}(2013)\citenamefont {Tamura},
  \citenamefont {Terada}, \citenamefont {Yamashita},\ and\ \citenamefont
  {Asama}}]{Tamura2013}%
  \BibitemOpen
  \bibfield  {author} {\bibinfo {author} {\bibfnamefont {Y.}~\bibnamefont
  {Tamura}}, \bibinfo {author} {\bibfnamefont {Y.}~\bibnamefont {Terada}},
  \bibinfo {author} {\bibfnamefont {A.}~\bibnamefont {Yamashita}}, \ and\
  \bibinfo {author} {\bibfnamefont {H.}~\bibnamefont {Asama}},\ }\href
  {\doibase 10.5772/56668} {\bibfield  {journal} {\bibinfo  {journal} {Int. J.
  Adv. Robot. Syst.}\ }\textbf {\bibinfo {volume} {10}} (\bibinfo {year}
  {2013}),\ 10.5772/56668}\BibitemShut {NoStop}%
\bibitem [{\citenamefont {Seitz}\ \emph {et~al.}(2016)\citenamefont {Seitz},
  \citenamefont {Seer}, \citenamefont {Klettner}, \citenamefont {Handel},\ and\
  \citenamefont {K{\"o}ster}}]{seitzTGF15}%
  \BibitemOpen
  \bibfield  {author} {\bibinfo {author} {\bibfnamefont {M.}~\bibnamefont
  {Seitz}}, \bibinfo {author} {\bibfnamefont {S.}~\bibnamefont {Seer}},
  \bibinfo {author} {\bibfnamefont {S.}~\bibnamefont {Klettner}}, \bibinfo
  {author} {\bibfnamefont {O.}~\bibnamefont {Handel}}, \ and\ \bibinfo {author}
  {\bibfnamefont {G.}~\bibnamefont {K{\"o}ster}},\ }in\ \href@noop {} {\emph
  {\bibinfo {booktitle} {Traffic and Granular Flows '15}}},\ \bibinfo {editor}
  {edited by\ \bibinfo {editor} {\bibfnamefont {W.}~\bibnamefont {Daamen}}\
  and\ \bibinfo {editor} {\bibfnamefont {V.}~\bibnamefont {Knoop}}}\ (\bibinfo
  {publisher} {Springer},\ \bibinfo {year} {2016})\BibitemShut {NoStop}%
\bibitem [{\citenamefont {Kosiantis}\ \emph {et~al.}(2007)\citenamefont
  {Kosiantis}, \citenamefont {Zaharakis},\ and\ \citenamefont
  {Pintelas}}]{kotsiantis2007supervised}%
  \BibitemOpen
  \bibfield  {author} {\bibinfo {author} {\bibfnamefont {S.}~\bibnamefont
  {Kosiantis}}, \bibinfo {author} {\bibfnamefont {I.}~\bibnamefont
  {Zaharakis}}, \ and\ \bibinfo {author} {\bibfnamefont {P.}~\bibnamefont
  {Pintelas}},\ }in\ \href@noop {} {\emph {\bibinfo {booktitle} {Emerging
  artificial intelligence applications in computer engineering}}},\ Vol.\
  \bibinfo {volume} {160}\ (\bibinfo {year} {2007})\ pp.\ \bibinfo {pages}
  {3--24}\BibitemShut {NoStop}%
\bibitem [{\citenamefont {{Microsoft Corp.}}(2012)}]{Kinect}%
  \BibitemOpen
  \bibfield  {author} {\bibinfo {author} {\bibnamefont {{Microsoft Corp.}}},\
  }\href@noop {} {\enquote {\bibinfo {title} {Kinect for {X}box 360},}\ }
  (\bibinfo {year} {2012}),\ \bibinfo {note} {{R}edmond, WA, USA.}\BibitemShut
  {Stop}%
\bibitem [{\citenamefont {Coffey}\ \emph {et~al.}(2004)\citenamefont {Coffey},
  \citenamefont {Kalmykov},\ and\ \citenamefont
  {Waldron}}]{coffey2004applications}%
  \BibitemOpen
  \bibfield  {author} {\bibinfo {author} {\bibfnamefont {W.~T.}\ \bibnamefont
  {Coffey}}, \bibinfo {author} {\bibfnamefont {Y.~P.}\ \bibnamefont
  {Kalmykov}}, \ and\ \bibinfo {author} {\bibfnamefont {J.~T.}\ \bibnamefont
  {Waldron}},\ }\href@noop {} {\emph {\bibinfo {title} {The Langevin Equation:
  with Applications to Stochastic Problems in Physics, Chemistry and Electrical
  Engineering}}}\ (\bibinfo  {publisher} {World Scientific},\ \bibinfo {year}
  {2004})\BibitemShut {NoStop}%
\bibitem [{\citenamefont {Corbetta}\ \emph {et~al.}(2014)\citenamefont
  {Corbetta}, \citenamefont {Bruno}, \citenamefont {Muntean},\ and\
  \citenamefont {Toschi}}]{corbetta2014TRP}%
  \BibitemOpen
  \bibfield  {author} {\bibinfo {author} {\bibfnamefont {A.}~\bibnamefont
  {Corbetta}}, \bibinfo {author} {\bibfnamefont {L.}~\bibnamefont {Bruno}},
  \bibinfo {author} {\bibfnamefont {A.}~\bibnamefont {Muntean}}, \ and\
  \bibinfo {author} {\bibfnamefont {F.}~\bibnamefont {Toschi}},\ }\href
  {\doibase 10.1016/j.trpro.2014.09.013} {\bibfield  {journal} {\bibinfo
  {journal} {Transportation Research Procedia}\ }\textbf {\bibinfo {volume}
  {2}},\ \bibinfo {pages} {96} (\bibinfo {year} {2014})}\BibitemShut {NoStop}%
\bibitem [{\citenamefont {Corbetta}\ \emph
  {et~al.}(2016{\natexlab{b}})\citenamefont {Corbetta}, \citenamefont {Lee},
  \citenamefont {Muntean},\ and\ \citenamefont {Toschi}}]{corbettaTGF15}%
  \BibitemOpen
  \bibfield  {author} {\bibinfo {author} {\bibfnamefont {A.}~\bibnamefont
  {Corbetta}}, \bibinfo {author} {\bibfnamefont {C.}~\bibnamefont {Lee}},
  \bibinfo {author} {\bibfnamefont {A.}~\bibnamefont {Muntean}}, \ and\
  \bibinfo {author} {\bibfnamefont {F.}~\bibnamefont {Toschi}},\ }in\ \href
  {\doibase 10.1007/978-3-319-33482-0\_7} {\emph {\bibinfo {booktitle} {Traffic
  and Granular Flows '15}}},\ \bibinfo {editor} {edited by\ \bibinfo {editor}
  {\bibfnamefont {W.}~\bibnamefont {Daamen}}\ and\ \bibinfo {editor}
  {\bibfnamefont {V.}~\bibnamefont {Knoop}}}\ (\bibinfo  {publisher}
  {Springer},\ \bibinfo {year} {2016})\ Chap.~\bibinfo {chapter}
  {7}\BibitemShut {NoStop}%
\bibitem [{\citenamefont {Seer}\ \emph {et~al.}(2014)\citenamefont {Seer},
  \citenamefont {Br{\"a}ndle},\ and\ \citenamefont {Ratti}}]{seer2014kinects}%
  \BibitemOpen
  \bibfield  {author} {\bibinfo {author} {\bibfnamefont {S.}~\bibnamefont
  {Seer}}, \bibinfo {author} {\bibfnamefont {N.}~\bibnamefont {Br{\"a}ndle}}, \
  and\ \bibinfo {author} {\bibfnamefont {C.}~\bibnamefont {Ratti}},\ }\href
  {\doibase 10.1016/j.trc.2014.08.012} {\bibfield  {journal} {\bibinfo
  {journal} {Transport. Res. C-Emer.}\ }\textbf {\bibinfo {volume} {48}},\
  \bibinfo {pages} {212} (\bibinfo {year} {2014})}\BibitemShut {NoStop}%
\bibitem [{\citenamefont {Br\v{s}\v{c}i\'{c}}\ \emph
  {et~al.}(2013)\citenamefont {Br\v{s}\v{c}i\'{c}}, \citenamefont {Kanda},
  \citenamefont {Ikeda},\ and\ \citenamefont {Miyashita}}]{brscic2013person}%
  \BibitemOpen
  \bibfield  {author} {\bibinfo {author} {\bibfnamefont {D.}~\bibnamefont
  {Br\v{s}\v{c}i\'{c}}}, \bibinfo {author} {\bibfnamefont {T.}~\bibnamefont
  {Kanda}}, \bibinfo {author} {\bibfnamefont {T.}~\bibnamefont {Ikeda}}, \ and\
  \bibinfo {author} {\bibfnamefont {T.}~\bibnamefont {Miyashita}},\ }\href
  {\doibase 10.1109/THMS.2013.2283945} {\bibfield  {journal} {\bibinfo
  {journal} {IEEE Trans. Human-Mach. Syst.}\ }\textbf {\bibinfo {volume}
  {43}},\ \bibinfo {pages} {522} (\bibinfo {year} {2013})}\BibitemShut
  {NoStop}%
\bibitem [{\citenamefont {Corbetta}\ \emph
  {et~al.}(2017{\natexlab{b}})\citenamefont {Corbetta}, \citenamefont {min
  Lee}, \citenamefont {Muntean},\ and\ \citenamefont
  {Toschi}}]{corbetta-colldyn17}%
  \BibitemOpen
  \bibfield  {author} {\bibinfo {author} {\bibfnamefont {A.}~\bibnamefont
  {Corbetta}}, \bibinfo {author} {\bibfnamefont {C.}~\bibnamefont {min Lee}},
  \bibinfo {author} {\bibfnamefont {A.}~\bibnamefont {Muntean}}, \ and\
  \bibinfo {author} {\bibfnamefont {F.}~\bibnamefont {Toschi}},\ }\href
  {\doibase 10.17815/CD.2017.10} {\bibfield  {journal} {\bibinfo  {journal}
  {Collective Dynamics}\ }\textbf {\bibinfo {volume} {1}},\ \bibinfo {pages}
  {1} (\bibinfo {year} {2017}{\natexlab{b}})}\BibitemShut {NoStop}%
\bibitem [{\citenamefont {Helbing}\ and\ \citenamefont
  {Johansson}(2009)}]{helbing2013pedestrian}%
  \BibitemOpen
  \bibfield  {author} {\bibinfo {author} {\bibfnamefont {D.}~\bibnamefont
  {Helbing}}\ and\ \bibinfo {author} {\bibfnamefont {A.}~\bibnamefont
  {Johansson}},\ }in\ \href@noop {} {\emph {\bibinfo {booktitle} {Encyclopedia
  of {C}omplexity and {S}ystems {S}cience}}},\ Vol.~\bibinfo {volume} {16},\
  \bibinfo {editor} {edited by\ \bibinfo {editor} {\bibfnamefont {R.~A.}\
  \bibnamefont {Meyers}}}\ (\bibinfo  {publisher} {Springer New York},\
  \bibinfo {year} {2009})\ pp.\ \bibinfo {pages} {6476--6495}\BibitemShut
  {NoStop}%
\bibitem [{\citenamefont {Ballerini}\ \emph {et~al.}(2008)\citenamefont
  {Ballerini}, \citenamefont {Cabibbo}, \citenamefont {Candelier},
  \citenamefont {Cavagna}, \citenamefont {Cisbani}, \citenamefont {Giardina},
  \citenamefont {Lecomte}, \citenamefont {Orlandi}, \citenamefont {Parisi},
  \citenamefont {Procaccini}, \citenamefont {Viale},\ and\ \citenamefont
  {Zdravkovic}}]{ballerini2008interaction}%
  \BibitemOpen
  \bibfield  {author} {\bibinfo {author} {\bibfnamefont {M.}~\bibnamefont
  {Ballerini}}, \bibinfo {author} {\bibfnamefont {N.}~\bibnamefont {Cabibbo}},
  \bibinfo {author} {\bibfnamefont {R.}~\bibnamefont {Candelier}}, \bibinfo
  {author} {\bibfnamefont {A.}~\bibnamefont {Cavagna}}, \bibinfo {author}
  {\bibfnamefont {E.}~\bibnamefont {Cisbani}}, \bibinfo {author} {\bibfnamefont
  {I.}~\bibnamefont {Giardina}}, \bibinfo {author} {\bibfnamefont
  {V.}~\bibnamefont {Lecomte}}, \bibinfo {author} {\bibfnamefont
  {A.}~\bibnamefont {Orlandi}}, \bibinfo {author} {\bibfnamefont
  {G.}~\bibnamefont {Parisi}}, \bibinfo {author} {\bibfnamefont
  {A.}~\bibnamefont {Procaccini}}, \bibinfo {author} {\bibfnamefont
  {M.}~\bibnamefont {Viale}}, \ and\ \bibinfo {author} {\bibfnamefont
  {V.}~\bibnamefont {Zdravkovic}},\ }\href@noop {} {\bibfield  {journal}
  {\bibinfo  {journal} {Proc. Natl. Acad. Sci.}\ }\textbf {\bibinfo {volume}
  {105}},\ \bibinfo {pages} {1232} (\bibinfo {year} {2008})}\BibitemShut
  {NoStop}%
\bibitem [{\citenamefont {Arechavaleta}\ \emph {et~al.}(2008)\citenamefont
  {Arechavaleta}, \citenamefont {Laumond}, \citenamefont {Hicheur},\ and\
  \citenamefont {Berthoz}}]{arechavaleta2008optimality}%
  \BibitemOpen
  \bibfield  {author} {\bibinfo {author} {\bibfnamefont {G.}~\bibnamefont
  {Arechavaleta}}, \bibinfo {author} {\bibfnamefont {J.}~\bibnamefont
  {Laumond}}, \bibinfo {author} {\bibfnamefont {H.}~\bibnamefont {Hicheur}}, \
  and\ \bibinfo {author} {\bibfnamefont {A.}~\bibnamefont {Berthoz}},\
  }\href@noop {} {\bibfield  {journal} {\bibinfo  {journal} {IEEE Trans.
  Robot}\ }\textbf {\bibinfo {volume} {24}},\ \bibinfo {pages} {5} (\bibinfo
  {year} {2008})}\BibitemShut {NoStop}%
\bibitem [{\citenamefont {Parisi}\ \emph {et~al.}(2009)\citenamefont {Parisi},
  \citenamefont {Gilman},\ and\ \citenamefont
  {Moldovan}}]{parisi2009modification}%
  \BibitemOpen
  \bibfield  {author} {\bibinfo {author} {\bibfnamefont {D.~R.}\ \bibnamefont
  {Parisi}}, \bibinfo {author} {\bibfnamefont {M.}~\bibnamefont {Gilman}}, \
  and\ \bibinfo {author} {\bibfnamefont {H.}~\bibnamefont {Moldovan}},\
  }\href@noop {} {\bibfield  {journal} {\bibinfo  {journal} {Phys. A}\ }\textbf
  {\bibinfo {volume} {388}},\ \bibinfo {pages} {3600} (\bibinfo {year}
  {2009})}\BibitemShut {NoStop}%
\bibitem [{\citenamefont {Corbetta}(2016)}]{corbetta2016multiscale}%
  \BibitemOpen
  \bibfield  {author} {\bibinfo {author} {\bibfnamefont {A.}~\bibnamefont
  {Corbetta}},\ }\emph {\bibinfo {title} {Multiscale crowd dynamics: physical
  analysis, modeling and applications}},\ \href@noop {} {Ph.D. thesis},\
  \bibinfo  {school} {Technische Universiteit Eindhoven} (\bibinfo {year}
  {2016})\BibitemShut {NoStop}%
\bibitem [{\citenamefont {Corbetta}\ \emph {et~al.}(2015)\citenamefont
  {Corbetta}, \citenamefont {Muntean},\ and\ \citenamefont
  {Vafayi}}]{corbetta2015MBE}%
  \BibitemOpen
  \bibfield  {author} {\bibinfo {author} {\bibfnamefont {A.}~\bibnamefont
  {Corbetta}}, \bibinfo {author} {\bibfnamefont {A.}~\bibnamefont {Muntean}}, \
  and\ \bibinfo {author} {\bibfnamefont {K.}~\bibnamefont {Vafayi}},\ }\href
  {\doibase 10.3934/mbe.2015.12.337} {\bibfield  {journal} {\bibinfo  {journal}
  {Math. Biosci. Eng.}\ }\textbf {\bibinfo {volume} {12}},\ \bibinfo {pages}
  {337} (\bibinfo {year} {2015})}\BibitemShut {NoStop}%
\bibitem [{\citenamefont {Willneff}(2003)}]{willneff2003spatio}%
  \BibitemOpen
  \bibfield  {author} {\bibinfo {author} {\bibfnamefont {J.}~\bibnamefont
  {Willneff}},\ }\emph {\bibinfo {title} {A {S}patio-{T}emporal {M}atching
  {A}lgorithm for 3{D} {P}article {T}racking {V}elocimetry}},\ \href@noop {}
  {Ph.D. thesis},\ \bibinfo  {school} {Institut f{\"u}r Geod{\"a}sie und
  Photogrammetrie an der Eidgen{\"o}ssichen Technischen Hochschule} (\bibinfo
  {year} {2003})\BibitemShut {NoStop}%
\bibitem [{\citenamefont {{The {OpenPTV} Consortium}}(12  )}]{OpenPTV}%
  \BibitemOpen
  \bibfield  {author} {\bibinfo {author} {\bibnamefont {{The {OpenPTV}
  Consortium}}},\ }\href@noop {} {\enquote {\bibinfo {title} {{OpenPTV: Open
  source particle tracking velocimetry}},}\ } (\bibinfo {year}
  {2012--})\BibitemShut {NoStop}%
\bibitem [{\citenamefont {Savitzky}\ and\ \citenamefont
  {Golay}(1964)}]{savitzky1964smoothing}%
  \BibitemOpen
  \bibfield  {author} {\bibinfo {author} {\bibfnamefont {A.}~\bibnamefont
  {Savitzky}}\ and\ \bibinfo {author} {\bibfnamefont {M.~J.}\ \bibnamefont
  {Golay}},\ }\href@noop {} {\bibfield  {journal} {\bibinfo  {journal}
  {Analytical chemistry}\ }\textbf {\bibinfo {volume} {36}},\ \bibinfo {pages}
  {1627} (\bibinfo {year} {1964})}\BibitemShut {NoStop}%
\bibitem [{\citenamefont {Gülan}\ \emph {et~al.}(2012)\citenamefont {Gülan},
  \citenamefont {L{\"u}thi}, \citenamefont {Holzner}, \citenamefont {Liberzon},
  \citenamefont {Tsinober},\ and\ \citenamefont {Kinzelbach}}]{PTV-liberzon}%
  \BibitemOpen
  \bibfield  {author} {\bibinfo {author} {\bibfnamefont {U.}~\bibnamefont
  {Gülan}}, \bibinfo {author} {\bibfnamefont {B.}~\bibnamefont {L{\"u}thi}},
  \bibinfo {author} {\bibfnamefont {M.}~\bibnamefont {Holzner}}, \bibinfo
  {author} {\bibfnamefont {A.}~\bibnamefont {Liberzon}}, \bibinfo {author}
  {\bibfnamefont {A.}~\bibnamefont {Tsinober}}, \ and\ \bibinfo {author}
  {\bibfnamefont {W.}~\bibnamefont {Kinzelbach}},\ }\href {\doibase
  10.1007/s00348-012-1371-8} {\bibfield  {journal} {\bibinfo  {journal}
  {Experiments in Fluids}\ }\textbf {\bibinfo {volume} {53}},\ \bibinfo {pages}
  {1469} (\bibinfo {year} {2012})}\BibitemShut {NoStop}%
\end{thebibliography}%

\end{document}